\documentstyle[11pt]{article}
\begin{document}
{\setlength{\oddsidemargin}{1.2in}
\setlength{\evensidemargin}{1.2in} }
\baselineskip 0.55cm
\title{Non-stationary rotating black holes: Entropy and
Hawking's radiation}

\author
{Ng. Ibohal\ddag\ and L. Dorendro\dag\,\\
Department of Mathematics, Manipur University,\\
Imphal 795003, Manipur, INDIA\\
\ddag\ E-mail: (i) ngibohal@iucaa.ernet.in,
(ii) ngibohal@rediffmail.com \\
\dag\ E-mail: dorendro@iucaa.ernet.in}
\date{December 30, 2004}
\maketitle

\begin{abstract}
In this paper we derive a class of {\sl non-stationary} rotating
solutions including Vaidya-Bonnor-de Sitter,
Vaidya-Bonnor-monopole and Vaidya-Bonnor-Kerr. The rotating
Viadya-Bonnor-de Sitter solution describes an embedded black hole
that the rotating Vaidya-Bonnor black hole is embedded into the
rotating de Sitter cosmological universe. In the case of the
Vaidya-Bonnor-Kerr, the rotating Vaidya-Bonnor solution is
embedded into the vacuum Kerr solution, and similarly,
Vaidya-Bonnor-monopole. By considering the charge to be function
of $u$ and $r$, we discuss the Hawking's evaporation of the masses
of variable-charged {\sl non-embedded}, non-rotating and rotating
Vaidya-Bonnor, and {\sl embedded} rotating, Vaidya-Bonnor-de
Sitter, Vaidya-Bonnor monopole and Vaidya-Bonnor-Kerr, black
holes. It is found that every electrical radiation of
variable-charged black holes will produce a change in the mass of
the body without affecting the Maxwell scalar in {\sl non-embedded
cases}; whereas in {\sl embedded cases} the Maxwell scalar, the
cosmological constant, monopole charge and the Kerr mass are not
affected by the radiation process. It is also found that during
the Hawking's radiation process, after the complete evaporation of
masses of these variable-charged black holes, the electrical
radiation will continue creating (i) negative mass naked
singularities in {\sl non-embedded} ones, and (ii) embedded
negative mass naked singularities in {\sl embedded} black holes.
The surface gravity, entropy and angular velocity of the horizon
are presented for each of these non-stationary black holes.  
Vaidya-Bonnor-Kerr black holes.

\end{abstract}

\begin{center}
{\bf 1. Introduction}
\end{center}
\setcounter{equation}{0}
\renewcommand{\theequation}{1.\arabic{equation}}

The Hawking's radiation [1] suggests that black holes which are
formed by collapse, are not completely black, but emit radiation
with a thermal spectrum. It means that, as the radiation carries
away energy, the black holes must presumably lose mass and
eventually disappear [2]. In an introductory survey Hawking and
Israel [3] have discussed the black hole radiation in three
possible ways with creative remarks --`So far there is no good
theoretical frame work with which to treat the final stages of a
black hole but there seem to be three possibilities: (i) The black
hole might disappear completely, leaving just the thermal
radiation that it emitted during its evaporation. (ii) It might
leave behind a non-radiating black hole of about the Planck mass.
(iii) The emission of energy might continue indefinitely creating
a negative mass naked singularity'. In [4] these three
possibilities of black hole radiation could be expressed in
classical spacetime metrics, by considering the charge $e$ of the
electromagnetic field to be function of the radial coordinate $r$
of {\sl non-stationary} Reissner-Nordstrom as well as Kerr-Newman
black holes. The variable-charge $e(r)$ with respect to the
coordinate $r$ is followed from Boulware's suggestion [5] that
{\it the stress-energy tensor may be used to calculate the change
in the mass due to the radiation}. According to Boulware's
suggestion, the energy momentum tensor of a particular black hole
can be used to calculate the change in the mass in order to
incorporate the Hawking's radiation effects in classical spacetime
metrics. This idea suggests to consider the stress-energy tensor
of electromagnetic field of different forms or functions from
those of {\sl stationary} Reissner-Nordstrom, as well as
Kerr-Newman, black holes as these two black holes do not seem to
have any direct Hawking's radiation effects. Thus, a variable
charge in the field equations will have the different function of
the energy momentum tensor of the charged black hole. {\sl Such a
variable charge $e$ with respect to the coordinate $r$ in
Einstein's equations is referred to as an electrical radiation
({\rm or} Hawking's electrical radiation) of the black hole}.
Every electrical radiation e(r) of the non-rotating as well as
rotating black holes leads to a reduction in its mass by some
quantity. If one considers such electrical radiation taking place
continuously for a long time, then a continuous reduction of the
mass will take place in the black hole body whether rotating or
non-rotating, and the original mass of the black hole will
evaporate completely. At that stage the complete evaporation will
lead the gravity of the object depending only on the
electromagnetic field, and not on the mass. We refer to such an
object with zero mass as an `instantaneous' naked singularity - a
naked singularity that exists for an instant and then continues
its electrical radiation to create negative mass. So this naked
singularity is different from the one mentioned in Steinmular {\it
et al.} [6], Tipler {\it et al.} [7] in the sense that an
`instantaneous' naked singularity, discussed in [6, 7] exists only
for an instant and then disappears.

It is emphasized that the time taken between two consecutive
radiation is supposed to be so short that one may not physically
realize how quickly radiation take place. Thus, it seems natural
to expect the existence of an `instantaneous' naked singularity
with zero mass only for an instant before continuing its next
radiation to create a negative mass naked singularity. This
suggests that it may also be possible in the common theory of
black holes that, as a black hole is invisible in nature, one may
not know whether, in the universe, a particular black hole has
mass or not, but electrical radiation may be detected on the black
hole surface. Immediately after the complete evaporation of the
mass, if one continues to radiate the remaining remnant, there
will be a formation of a new mass. If one repeats the electrical
radiation further, the new mass will increase gradually and then
the spacetime geometry will represent the `negative mass naked
singularity'. In order to study Hawking's radiation in classical
spacetime metrics, the Boulware's suggestion leads us to consider
the stress-energy tensors of electromagnetic field of different
forms or functions from those of Reissner-Nordstrom, as well as
Kerr-Newman, black holes as these two black holes do not seem to
have any direct Hawking's radiation effects. Thus, (i) the changes
in the mass of black holes, (ii) the formation of `instantaneous'
naked singularities with zero mass and (iii) the creation of
`negative mass naked singularities' in {\sl stationary}
Reissner-Nordstrom as well as Kerr-Newman black holes [4] may
presumably  be the correct formulations in classical spacetime
metrics of the three possibilities of black hole evaporation
suggested by Hawking and Israel [3].

The aim of this paper is to study the relativistic aspect of
Hawking's radiation in {\sl non-stationary}, embedded and
non-embedded black holes by considering the charge $e(u)$ to be
variable with respect to the coordinate $r$, i.e. $e(u,r)$. This
consideration of the variable charge $e(u,r)$ will lead to a
different energy momentum tensor than the original ones of both
the {\sl non-stationary}, embedded and non-embedded, black holes.
Here, according to Cai, et. al. [8], the embedded black hole means
that the rotating Vaidya-Bonnor black hole is embedded into the
rotating de Sitter cosmological universe to produce the rotating
embedded Vaidya-Bonnor-de Sitter black hole and so on. It is also
noted that all the black holes which are extended from the
Vaidya-Bonnor solutions, are all {\sl non-stationary} black holes.
For examples, the rotating Vaidya-Bonnor, rotating
Vaidya-Bonnor-de Sitter, rotating Vaidya-Bonnor-monopole and
rotating Vaidya-Bonnor-Kerr black holes, which are to be discussed
later in this paper, are all {\sl non-stationary} black holes.
Here, using Wang-Wu functions, it is shown the derivation of
embedded rotating black holes, Vaidya-Bonnor-de Sitter,
Vaidya-Bonnor-monopole and Vaidya-Bonnor-Kerr in a simple
analytic method. Then the relativistic aspect
of Hawking radiation effects has been studied in these {\sl
non-stationary} black holes mentioned above. The results are
summarized in the form of theorems as follows:

\newtheorem{theorem}{Theorem}
\begin{theorem}
Every electrical radiation of {\sl non-stationary} embedded and
non-embedded black holes will produce a change in the mass of the
bodies without affecting the Maxwell scalar.
\end{theorem}
\begin{theorem}
The non-rotating and rotating Vaidya-Bonnor black holes will lead
to the formation of instantaneous naked singularities with zero
mass during the Hawking's evaporation process of electrical
radiation.
\end{theorem}
\begin{theorem}
The non-stationary rotating embedded black holes will lead to the
formation of instantaneous charged black holes, with accord to the
nature of the back ground spaces, during the Hawking's radiation
process.
\end{theorem}
\begin{theorem}
During the radiation process, after the complete evaporation of
masses of both variable-charged non-rotating and rotating
Vaidya-Bonnor black holes, the electrical radiation will continue
indefinitely creating negative mass naked singularities.
\end{theorem}
\begin{theorem}
During the radiation process, after the complete evaporation of
masses of variable-charged non-stationary embedded black holes,
the electrical radiation will continue indefinitely creating
embedded negative mass naked singularities.
\end{theorem}
\begin{theorem}
If an electrically radiating non-stationary black hole is embedded
into a space, it will continue to embed into the same space
forever.
\end{theorem}
\begin{theorem}
Every embedded black hole, stationary or non-stationary, is expressible
in Kerr-Schild ansatz.
\end{theorem}

It is found that the theorems 1, 2 and 4 are in favour of the
first, second and third possibilities of the suggestions made by
Hawking and Israel [3]. But theorem 4 provides a violation of
Penrose's cosmic-censorship hypothesis that `no naked singularity
can ever be created' [9]. Theorems 3, 5 and 6 show the various
stages of the life of embedded non-stationary radiating black
holes. Theorem 7 shows that every embedded black hole is a
solution of Einstein's field equations.

Here, it is more appropriate to use the phrase `change in the
mass' rather then `loss of mass' as there is a possibility of
creating new mass after the exhaustion of the original mass, if
one repeats the same process of electrical radiation. This can be
seen latter in this paper. Hawking's radiation is being
incorporated, in the classical general relativity describing the
change in mass appearing in the classical space-time metrics. In
section 2 we derive a class of rotating {\sl non-stationary}
solutions, including Vaidya-Bonnor-de Sitter,
Vaidya-Bonnor-monopole and Vaidya-Bonnor-Kerr describing the
embedded black holes. The surface gravity, entropy and angular
velocity of the horizons are presented as these are the three
important properties of black holes. In section 3 we study the
relativistic aspect of Hawking's radiation in {\sl
non-stationary}, embedded and non-embedded black holes by
considering the charge $e(u)$ to be variable with respect to the
coordinate $r$, i.e. $e(u,r)$. We conclude with the remarks and
suggestions of our results presented in this paper in section 4.
In an appendix we cite the NP quantities for a rotating metric
with functions of two variables. We also present the general
formula for the surface gravity and entropy on a horizon of a
black hole. We use the differential form structure in Newman and
Penrose (NP) formalism [10] developed by McIntosh and Hickman [11]
in (--2) signature.

\begin{center}
{\bf 2. Rotating non-stationary solutions}
\end{center}
\setcounter{equation}{0}
\renewcommand{\theequation}{2.\arabic{equation}}

In this section we shall derive a class of {\sl non-stationary}
rotating solutions describing embedded black holes in genaral
relativity, by using the Wang-Wu functions $q_n$ [12] given in
(A11) for a spherically symmetric metric with functions of two
variables (A1) [13].
 \vspace*{.15in}

{\sl (i) Rotating Vaidya-Bonnor-de Sitter solution}

\vspace*{.15in} We shall combine the rotating Vaidya-Bonnor
solution with the rotating de Sitter solution [13], if the Wang-Wu
functions $q_n(u)$ are chosen as
\begin{eqnarray}
\begin{array}{cc}
q_n(u)=&\left\{\begin{array}{ll}
M(u),&{\rm when }\;\;n=0\\
-e^2(u)/2, &{\rm when }\;\;n=-1\\
\Lambda^*/6, &{\rm when}\;\;n=3\\
0, &{\rm when }\;\;n\neq 0, -1, 3
\end{array}\right.
\end{array}
\end{eqnarray}
where $M(u)$ and $e(u)$ are related with the mass and the charge
of rotating Vaidya-Bonnor solution. Thus, using this $q_n(u)$
in (A11) we obtain the mass function as
\begin{equation}
M(u,r)=M(u)-{e^2(u)\over 2r}+{\Lambda^*\,r^3\over 6}.
\end{equation}
The line element describing a rotating Vaidya-Bonnor-de Sitter
solution takes the form
\begin{eqnarray}
d s^2&=&\Big[1-R^{-2}\Big\{2rM(u)+{\Lambda^*\,r^4\over
3}-e^2(u)\Big\}\Big]\,du^2+2du\,dr \cr
&&+2aR^{-2}\Big\{2rM(u)+{\Lambda^*\,r^4\over 3}
-e^2(u)\Big\}\,{\rm sin}^2 \theta\,du\,d\phi \cr && -2a\,{\rm
sin}^2\theta\,dr\,d\phi -R^2d\theta^2 \cr && -\Big\{(r^2+a^2)^2
-\Delta^*a^2\,{\rm sin}^2\theta\Big\}\,R^{-2}{\rm
sin}^2\theta\,d\phi^2,
\end{eqnarray}
where $\Delta^*=r^2-2rM(u)-{\Lambda^*\,r^4/3} +a^2+e^2(u)$. Here,
$a$ is the  non-zero rotational parameter per unit mass.
Then the other quantities for the metric are
\begin{eqnarray}
\rho^* &=& {1\over K\,R^2\,R^2}\Big\{e^2(u) +
\Lambda^*r^4\Big\},\cr
 p &=& {1\over K\,R^2\,R^2}\Big\{e^2(u) -
\Lambda^*r^2(r^2 +2a^2{\rm cos}\theta)\Big\},\cr
\mu^* &=&-{1\over
K\,R^2\,R^2}\Big[2\,r^2\Big\{M(u)_{,u}-{1\over
r}e(u)\,e(u)_{,u}\Big\} \cr && +a^2{\rm
sin}^2\theta\,\Big\{M(u)_{,u}-{1\over
r}e(u)\,e(u)_{,u}\Big\}_{,u}\Big],\cr
\omega &=&{-i\,a\,{\rm sin}\,\theta\,\over{\surd
2\,K\,R^2R^2}}\Big\{R\,M(u)_{,u}-2e(u)\,e(u)_{,u}\Big\},\\
\Lambda &=&{\Lambda^*\,r^2\over{6\,R^2}},\\
\psi_2&=&{1\over\overline R\,\overline R\,R^2}\Big[e^2(u)
-R\,M(u)+{\Lambda^*r^2\over 3}a^2{\rm cos}^2\theta\Big], \cr
\psi_3 &=&{-i\,a\,{\rm sin}\theta\over 2\surd 2\overline
R\,\overline R\,R^2} \Big[\Big(4\,r+\overline
R\Big)M(u)_{,u}\Big\}- 4e(u)\,e(u)_{,u}\Big], \cr
\psi_4&=&{{a^2\,{\rm sin}^2\theta}\over 2\overline R\,\overline
R\,R^2\,R^2}\,\Big[R^2\Big\{r\,M(u)_{,u} -e(u)e(u)_{,u}\Big\}_{,u}
\cr &&-2r\Big\{r\,M(u)_{,u} -e(u)e(u)_{,u}\Big\}\Big].
\end{eqnarray}
The metric (2.3) will describe a cosmological black holes with the
horizons at the values of $r$ for which $\Delta^*=0$ having four
roots $r_{+\,+}$, $r_{+\,-}$, $r_{-\,+}$ and $r_{-\,-}$ given in
appendix (A15) and (A16). The first three values will describe
respectively the event horizon, the Cauchy horizin and the
cosmological horizon. The surface gravity of the horizon at
$r=r_{+\,+}$ is
\begin{eqnarray}
{\cal
K}={-\Big[{1\over{r\,R^2}}\,\Big\{r\,\Big(r-M(u)-{\Lambda^*r^3\over
6 }\Big)+{e^2(u)\over 2}\Big\}\Big]}_{r=r_{+\,+}}.
\end{eqnarray}
Then the entropy and angular velocity of the horizon are
respectively found as,
\begin{eqnarray}
{\cal S}=\pi\,(r^2+a^2)_{r=r_{+\,+}},\;\;\; {\rm and}\;\;\;
\Omega_{\rm H}={a\,\{2\,r\,M(u)+(\Lambda^*r^4/3)
-e^2(u)\}\over{\{r^2+a^2}\}^2}\Big|_{r=r_{+\,+}}.
\end{eqnarray}
In this rotating solution (2.3), the Vaidya-Bonnor null fluid is
interacting with the non-null electromagnetic field on the de
Sitter cosmological space. Thus, the total energy momentum tensor
(EMT) for the rotating solution (2.3) takes the following form:
\begin{eqnarray}
T_{ab} = T^{(\rm n)}_{ab} +T^{(\rm E)}_{ab}+T^{(\rm C)}_{ab},
\end{eqnarray}
where the EMTs for the rotating null fluid, the electromagnetic
field and cosmological matter field are given respectively
\begin{eqnarray}
T^{(\rm n)}_{ab}&=& \mu^*\,\ell_a\,\ell_b +
2\,\omega\,\ell_{(a}\,\overline
m_{b)}+2\,\overline\omega\,\ell_{(a}\,m_{b)},\\
T^{(\rm E)}_{ab} &=&2\,\rho^{*(\rm E)}\{\ell_{(a}\,n_{b)}
+m_{(a}\overline m_{b)}\},\\
T^{(\rm C)}_{ab}&=&2\{\rho^{*(\rm C)}\,\ell_{(a}\,n_{b)} + p^{(\rm
C)}\,m_{(a}\overline m_{b)}\}.
\end{eqnarray}
where, $\mu^*$ and $\omega$ are given in (2.4) and
\begin{eqnarray*}
\rho^{*(\rm E)}&=& p^{(\rm E)}= {e^2(u)\over {K\,R^2\,R^2}}, \:\:
\rho^{*(\rm C)}= {\Lambda^*r^4\over {K\,R^2\,R^2}}, \cr  p^{(\rm
C)}&=& -{\Lambda^*r^2\over{K\,R^2\,R^2}} \Big(r^2+2a^2\,{\rm
cos}^2\theta\Big).
\end{eqnarray*}
Now, for future use we may, without loss of generality, have a
decomposition of the Ricci scalar $\Lambda$, given in (2.5), as
\begin{equation}
\Lambda={\Lambda^{(\rm E)}+\Lambda^{(\rm C)}},
\end{equation}
where $\Lambda^{(\rm E)}$ is the {\sl zero} Ricci scalar for the
electromagnetic field and $\Lambda^{(\rm C)}$ is the {\sl
non-zero} cosmological Ricci scalar with $\Lambda^{(\rm
C)}=(\Lambda^*\,r^2/6\,R^2)$. The appearance of $\omega$ shows
that the Vaidya-Bonnor null fluid is rotating as the expression of
$\omega$ in (2.4) involves the rotating parameter $a$ coupling
with $M(u)_{,u}$ both are non-zero quantities for a rotating
Vaidya-Bonnor null radiating universe. If we set $a=0$, we recover
the non-rotating Vaidya-Bonnor-de Sitter solution and then the
energy-momentum tensor (2.9) can be written in the form of Guth's
modification of $T_{ab}$ [14] as
\begin{equation}
T_{ab} =T^{(\rm n)}_{ab}+T^{(\rm E)}_{ab} + \Lambda^*g_{ab}
\end{equation}
where $T^{(\rm E)}_{ab}$ is the energy-momentum tensor for
non-null electromagnetic field and $g_{ab}$ is the non-rotating
Vaidya-Bonnor-de Sitter metric tensor. From this, without loss of
generality, the EMT (2.9) can be regarded as the extension of
Guth's modification of energy-momentum tensor in rotating spaces.
The rotating Vaidya-Bonnor-de Sitter metric can be written in
Kerr-Schild form:
\begin{equation}
g_{ab}^{\rm VBdS}=g_{ab}^{\rm dS} +2Q(u,r,\theta)\ell_a\ell_b
\end{equation}
where $Q(u,r,\theta) = -\{rM(u)-e^2(u)/2\}\,R^{-2}$. Here,
$g_{ab}^{\rm dS}$ is the rotating de Sitter metric and $\ell_a$ is
geodesic, shear free, expanding and non-zero twist null vector for
both $g_{ab}^{\rm dS}$ as well as $g_{ab}^{\rm VBdS}$ given in (A2)
with (A7). The above Kerr-Schild form can also be written on
the rotating Vaidya-Bonnor
background:
\begin{equation}
g_{ab}^{\rm VBdS}=g_{ab}^{\rm VB} +2Q(r,\theta)\ell_a\ell_b
\end{equation}
where $Q(r,\theta) =-(\Lambda^*r^4/6)\,R^{-2}$. These two
Kerr-Schild forms (2.15) and (2.16) prove the non-stationary
version of theorem 7 in the case of rotating Vaidya-Bonnor-de
Sitter solution. If we set $M(u)$ and $e(u)$ are both constant,
this Kerr-Schild form (2.16) will be that of Kerr-Newman-de Sitter
black hole. The rotating Vaidya-Bonnor-de Sitter metric will
describe a non-stationary spherically symmetric solution whose
Weyl curvature tensor is algebraically special in Petrov
classification possessing a geodesic, shear free, expanding and
non-zero twist null vector $\ell_a$ given in (A2) below. One can
easily recover a rotating Vaidya-de Sitter metric from this
Vaidya-Bonnor-de Sitter solution by setting the charge $e(u)=0$.
If one sets $a=0$, $e(u)=0$ in (2.3), one can also obtain the
standard non-rotating Vaidya-de Sitter solution [15]. Ghosh and
Dadhich [16] have studied the gravitational collapse problem in
non-rotating Vaidya-de Sitter space by identifying the de Sitter
cosmological constant $\Lambda^*$ with the bag constant of the
null strange quark fluid. Also if one sets $a=0$ in (2.3) one can
recover the non-rotating Vaidya-Bonnor-de Sitter black hole [17].
It certainly indicates that the embedded solution (2.3) can be
derived by using Wang-Wu functions (2.1) in the rotating metric
(A1). It is emphasized that the Vaidya-Bonnor-de Sitter solution,
when $M(u)$ and $e(u)$ are set constants, will recover the
Kerr-Newman-de Sitter black hole. We find that the Kerr-Newman-de
Sitter black hole, obtained with the constant parameters $M$
and $e$, is different from the one derived by Carter
[18], Mallet [19] and Xu [20] in terms of involving $\Lambda^{*}$.

\vspace*{.15in}

{\sl (ii) Non-stationary rotating Vaidya-Bonnor-monopole solution}

\vspace*{.15in}

 Here we shall study the rotating non stationary Vaidya-Bonnor
monopole solution by choosing the Wang-Wu functions $q_n(u)$ as
\begin{eqnarray}
\begin{array}{cc}
q_n(u)=&\left\{\begin{array}{ll}
M(u),&{\rm when }\;\;n=0\\
b/2, &{\rm when}\;\;n=1\\
-e^2(u)/2, &{\rm when }\;\;n=-1\\
0, &{\rm when }\;\;n\neq 0,1,-1
\end{array}\right.
\end{array}
\end{eqnarray}
where $M(u)$ and $e(u)$ are related with the mass and the charge
of rotating Vaidya-Bonnor solution. Thus, by using this $q_n(u)$,
we obtain the mass function as,
\begin{equation}
M(u,r)=M(u)+ \frac{rb}{2}-{e^2(u)\over 2r}.
\end{equation}
The line element describing a rotating Vaidya-Bonnor-monopole
solution take the form
\begin{eqnarray}
ds^2&=&[1-R^{-2}\{2rM(u)-e^2(u)+br^2\}]\,du^2+2du\,dr \cr
&+&2aR^{-2}\{2rM(u)-e^2(u)+br^2\}\,{\rm sin}^2
\theta\,du\,d\phi-2a\,{\rm sin}^2\theta\,dr\,d\phi \cr
&-&R^2d\theta^2-\{(r^2+a^2)^2 -\Delta^*a^2\,{\rm
sin}^2\theta\}\,R^{-2}{\rm sin}^2\theta\,d\phi^2,
\end{eqnarray}
where $\Delta^*=r^2(1-b)-2rM(u)+a^2+e^2(u)$. Here, $a$ is the
rotational parameter per unit mass, $b$ is the monopole
constant, $e(u)$ represents the charge of Vaidya-Bonnor solution
and $M(u)$ represents the mass function of rotating Vaidya-Bonnor
null radiating fluid. Then the other quantities for the metric are
\begin{eqnarray}
\rho^* &= &{1\over K\,R^2\,R^2}\Big\{e^2(u) + b\,r^2\Big\}, \cr
p&=& {1\over K\,R^2\,R^2}\Big\{e^2(u) - b\,a^2{\rm
cos}^2\theta\Big\},\cr \mu^*&=&-{1\over
K\,R^2\,R^2}\Big[2\,r\,M(u)_{,u}-{1\over
r}e(u)\,e(u)_{,u}\Big\},\cr &&+a^2{\rm
sin}^2\theta\,\Big\{M(u)_{,u}-{1\over
r}e(u)\,e(u)_{,u}\Big\}_{,u}\Big],\cr
\omega &=&{-i\,a\,{\rm sin}\,\theta\,\over{\surd
2\,K\,R^2R^2}}\Big\{R\,M(u)_{,u}-2e(u)\,e(u)_{,u}\Big\},\\
\Lambda&=& {b\over 12\,R^2},\\
\psi_2&=&{1\over{\overline R\,\overline R\,R^2}}\Big[- M(u)\,R +
e^2(u)-{b\over 6}\,\Big(RR+2\,r\,a\,i{\rm cos}\theta\Big)\Big],\cr
\psi_3 &=&{-i\,a\,{\rm sin}\theta\over 2\surd 2\overline
R\,\overline R\,R^2} \Big[\Big(4\,r+\overline
R\Big)M(u)_{,u}\Big\}- 4e(u)\,e(u)_{,u}\Big], \cr
\psi_4&=&{{a^2\,{\rm sin}^2\theta}\over 2\overline R\,\overline
R\,R^2\,R^2}\,\Big[R^2\Big\{r\,M(u)_{,u} -e(u)e(u)_{,u}\Big\}_{,u}
\cr &&-2r\Big\{r\,M(u)_{,u} -e(u)e(u)_{,u}\Big\}\Big].
\end{eqnarray}
The metric (2.19) will describe a black hole with
the horizons at the values of $r$ for which $\Delta^*=0$ having
two roots $r_{+}$ and $r_{-}$:
\begin{eqnarray}
r_{\pm}={1\over{1-b}}\,[M(u)\pm
\sqrt{M^2(u)-(1-b)\,\{a^2+e^2(u)\}}]
\end{eqnarray}
From this we observe that the value of $b$ must lie  in $0<b<1$,
with the horizon at $r=r_{+}$. Then the aurface gravity of the
horizon  at $r=r_+$ is
\begin{eqnarray}
{\cal
K}={-1\over{r_{+}\,R^2}}\,\Big[r_{+}\,\Big\{r_{+}\,\Big(r-{b\over
2}\Big)-M(u)\Big\}+{e^2(u)\over 2}\Big].
\end{eqnarray}
The entropy and angular velocity of the horizon are
respectively obtained as follows:
\begin{eqnarray}
{\cal S}=\pi\,(r^2_{+}+a^2),\;\;\; {\rm and}\;\;\; \Omega_{\rm H}
={a\,\{2\,r\,M(u)-e^2(u)+b\,r^2\}\over{\{r^2+a^2}\}^2}\Big|_{r=r_{+}}.
\end{eqnarray}
In this rotating solution (2.19), the matter field describing
monopole particles is interacting with the non-null
electromagnetic field. Thus, the total energy momentum tensor
(EMT) for the rotating solution (2.19) takes the following form:
\begin{eqnarray}
T_{ab} = T^{(\rm n)}_{ab}+T^{(\rm E)}_{ab}+T^{(\rm m)}_{ab},
\end{eqnarray}
where the EMTs for the monopole matter field and electromagnetic
field are given respectively
\begin{eqnarray}
T^{(\rm n)}_{ab}&=& \mu^*\,\ell_a\,\ell_b +
2\,\omega\,\ell_{(a}\,\overline
m_{b)}+2\,\overline\omega\,\ell_{(a}\,m_{b)},\\
T^{(\rm E)}_{ab} &=&2\,\rho^{*(\rm E)}\{\ell_{(a}\,n_{b)}
+m_{(a}\overline m_{b)}\},\\
T^{(\rm C)}_{ab}&=&2\{\rho^{*(\rm C)}\,\ell_{(a}\,n_{b)} + p^{(\rm
C)}\,m_{(a}\overline m_{b)}\}.
\end{eqnarray}
 where $\mu^*$ and $\omega$ are unchanged given in (2.20)
 and other quantities are
\begin{eqnarray}
\rho^{*(\rm E)}&=& p^{(\rm E)}= {e^2(u)\over {K\,R^2\,R^2}}, \:\:
\rho^{*(\rm m)}= {b\,r^2\over {K\,R^2\,R^2}}, \cr  p^{(\rm m)}&=&
{1\over{K\,R^2\,R^2}} \Big\{e^2(u)-ba^2\,{\rm cos}^2\theta\Big\}.
\end{eqnarray}
Now, for feature use we may, without loss of generality, have a
decomposition of the Ricci scalar $\Lambda$, given in (2.21), as
\begin{equation}
\Lambda={\Lambda^{(\rm E)}+\Lambda^{(\rm m)}},
\end{equation}
where $\Lambda^{(\rm E)}$ is the {\sl zero} Ricci scalar for the
electromagnetic field and $\Lambda^{(\rm m)}$ is the {\sl non-zero}
Ricci scalar for monopole field with $\Lambda^{(\rm
m)}={(b/{12\,R^2}})$. One has also seen the interaction of the
rotating parameter $a$ with the monopole constant $b$ in the
expression of $p^{(\rm m)}$, which makes difference between the
rotating as well as non-rotating monopole solutions. We also find
that the solution (2.19) with $M(u) = e(u) = 0$ represents
a rotating monopole solution, which is Petrov type $\rm D$ with the
Weyl scalar $\psi_2$ given in (2.22) whose repeated principal null
vector $\ell_{a}$ is shear free, rotating and non-zero twist. The
Vaidya-Bonnor-monopole metric can be written in Kerr-Schild form
\begin{equation}
g_{ab}^{\rm VBm}=g_{ab}^{\rm m} +2Q(u,r,\theta)\ell_a\ell_b
\end{equation}
where $Q(u,r,\theta) = -\{rM(u)-e^2(u)/2\}\,R^{-2}$. Here,
$g_{ab}^{\rm m}$ is the rotating monopole solution and $\ell_a$ is
geodesic, shear free, expanding and non-zero twist null vector for
both $g_{ab}^{\rm m}$ as well as $g_{ab}^{\rm VBm}$. The above
Kerr-Schild form can also be written on the rotating Vaidya-Bonnor
background as
\begin{equation}
g_{ab}^{\rm VBm}=g_{ab}^{\rm VB} +2Q(r,\theta)\ell_a\ell_b
\end{equation}
where $Q(r,\theta) =-(br^2/6)R^{-2}$. These two Kerr-Schild forms
(2.32) and (2.33) show the proof of the non-stationary version of
theorem 7 in
the case of rotating Vaidya-Bonnor-monopole solution. If we set
$M(u)$ and $e(u)$ are both constant, this Kerr-Schild form (2.32)
will be that of Kerr-Newman-monopole black hole. The rotating
Vaidya-Bonnor-monopole metric will describe a non-stationary
spherically symmetric solution whose Weyl curvature tensor is
algebraically special in Petrov classification possessing a
geodesic, shear free, expanding and non-zero twist null vector
$\ell_a$. One can easily recover a rotating Vaidya-monopole metric
from this Vaidya-Bonnor-monopole solution by setting the charge
$e(u)=0$. If one sets $a=0$, $e(u)=0$ in (2.19), one can also
obtain the standard non-rotating Vaidya-monopole solution [12].

\vspace*{.25in}

{\sl (iii) Non-stationary rotating Vaidya-Bonnor-Kerr solution}

\vspace*{.25in}

We shall combine the rotating Vaidya-Bonnor solution with the
rotating Ker solution, if the Wang-Wu functions $q_n(u)$ are
chosen such that
\begin{eqnarray}
\begin{array}{cc}
q_n(u)=&\left\{\begin{array}{ll}
\tilde{m}+M(u),&{\rm when }\;\;n=0\\
-e^2(u)/2, &{\rm when }\;\;n=-1\\
 0, &{\rm when }\;\;n\neq 0, -1

\end{array}\right.
\end{array}
\end{eqnarray}
where $\tilde{m}$ is the mass of Kerr black hole, and $M(u)$ and
$e(u)$ are related with the mass and the charge of rotating
Vaidya-Bonnor solution. Thus, using this $q_n(u)$ in (A11)
we obtain the mass function
\begin{equation}
M(u,r)=\tilde{m}+M(u)-{e^2(u)\over 2}.
\end{equation}
The line element representing a rotating Vaidya-Bonnor-Kerr
solution takes the form
\begin{eqnarray}
d
s^2&=&\Big[1-R^{-2}\Big\{2r\Big(\tilde{m}+M(u)\Big)
-e^2(u)\Big\}\Big]\,du^2+2du\,dr
\cr &&+2aR^{-2}\Big\{2r\Big(\tilde{m}+M(u)\Big)-e^2(u)\Big\}\,{\rm
sin}^2 \theta\,du\,d\phi \cr && -2a\,{\rm sin}^2\theta\,dr\,d\phi
-R^2d\theta^2 \cr && -\Big\{(r^2+a^2)^2 -\Delta^*a^2\,{\rm
sin}^2\theta\Big\}\,R^{-2}{\rm sin}^2\theta\,d\phi^2,
\end{eqnarray}
where $\Delta^*=r^2-2r{\tilde{m}+M(u)}+a^2+e^2(u)$. Here, $a$ is
the non-zero rotational parameter per unit mass. Then the other
quantities for the metric are
\begin{eqnarray}
\rho^* &=& {e^2(u)\over K\,R^2\,R^2},\;\;\;\; p= {e^(u)\over
K\,R^2\,R^2},\;\;\;\;
 \Lambda=0,\cr \psi_2&=&{-1\over\overline
R\,\overline
R\,R^2}\Big[R\,\Big\{\tilde{m}+M(u)\Big\}-e^2(u)\Big], \\
\psi_3 &=&{-i\,a\,{\rm sin}\theta\over 2\surd 2\overline
R\,\overline R\,R^2} \Big[\Big(4\,r+\overline R\Big)M(u)_{,u}-
4e(u)\,e(u)_{,u}\Big], \cr \psi_4&=&{{a^2\,{\rm sin}^2\theta}\over
2\overline R\,\overline R\,R^2\,R^2}\,\Big[R^2\Big\{r\,M(u)_{,u}
-e(u)e(u)_{,u}\Big\}_{,u} \cr &&-2r\Big\{r\,M(u)_{,u}
-e(u)e(u)_{,u}\Big\}\Big], \nonumber
\end{eqnarray}
and $\mu^*$ and $\omega$ are remained the same, given in case $2(ii)$.
The solution (2.36) will describe a black hole if
${\tilde{m}+M(u)}>{a^2+e^2(u)}$ with external horizon at
$r_{+}={\tilde{m}+M(u)}+\surd{[\{\tilde{m}+M(u)\}^2-\{a^2+e^2(u)\}]}$,
internal Cauchy horizon at
$r_{-}={\tilde{m}+M(u)}\\-\surd{[\{\tilde{m}+M(u)\}^2-\{a^2+e^2(u)\}]}$
and non stationary limit surface $r\equiv
r_{e}(u,\theta)\\={\tilde{m}+M(u)}+\surd{[\{\tilde{m}+M(u)\}^2-a^2\,{\rm
cos}^2\theta-e^2(u)]}$. The surface gravity of the event horizon
at $r=r_{+}$ is
\begin{eqnarray}
{\cal
K}=-\Big|\frac{1}{r\,R^2}\,\Big[r\,\sqrt{\{\tilde{m}+M(u)\}^2-a^2-e^2(u)}
+\frac{e^2(u)}{2}\Big]\Big|_{r=r_{+}}.
\end{eqnarray}
The entropy and angular velocity of the black hole at the
horizon are respectively found as
\begin{eqnarray}
&&{\cal S} =
2\pi\,\Big\{\tilde{m}+M(u)\Big\}\,\Big[{\tilde{m}+M(u)},
+\sqrt{\{\tilde{m}+M(u)\}^2-a^2-e^2(u)}\Big]-e^2(u),\cr
&&\Omega_{\rm H}
={a\,\Big\{2\,r\Big(\tilde{m}+M(u)\Big)-e^2(u)\Big\}\over(r^2+a^2)^2}\Big|_{r=r_{+}}.
\end{eqnarray}
In this rotating solution (2.36), the Vaidya null is interacting
with the non-null electromagnetic field. Thus, the total energy
momentum tensor (EMT) for the rotating solution (2.36) takes the
following form:
\begin{eqnarray}
T_{ab} = T^{(\rm n)}_{ab} +T^{(\rm E)}_{ab}
\end{eqnarray}
where the EMTs for the null radiating fluid field and electromagnetic
field are given respectively
\begin{eqnarray}
T^{(\rm n)}_{ab}&=& \mu^*\,\ell_a\,\ell_b +
2\,\omega\,\ell_{(a}\,\overline
m_{b)}+2\,\overline\omega\,\ell_{(a}\,m_{b)},\\
T^{(\rm E)}_{ab} &=&2\,\rho^{*(\rm E)}\{\ell_{(a}\,n_{b)}
+m_{(a}\overline m_{b)}\}.
\end{eqnarray}
The appearance of non-vanishing $\omega$ shows the null fluid is
rotating as the expression of omega involves the rotating
parameter $a$ coupling with $M_{,u}$, both non-zero quantities for
a rotating Vaidya-Bonnor null radiating universe. This rotating
Vaidya-Bonnor-Kerr metric can be written in Kerr-Schild form as
\begin{equation}
g_{ab}^{\rm VBK}=g_{ab}^{\rm K} +2Q(u,r,\theta)\ell_a\ell_b
\end{equation}
where $Q(u,r,\theta) = -\{rM(u)-e^2(u)/2\}R^{-2}$. Here, $g_{ab}^{\rm
K}$ is the rotating Kerr metric and $\ell_a$ is geodesic, shear
free, expanding and non-zero twist null vector for both
$g_{ab}^{\rm K}$ as well as $g_{ab}^{\rm VBK}$. The above Kerr-Schild
form can also be written on the rotating Vaidya-Bonnor background
\begin{equation}
g_{ab}^{\rm VBK}=g_{ab}^{\rm VB} +2Q(r,\theta)\ell_a\ell_b
\end{equation}
where $Q(r,\theta) = -r\tilde{m}\,R^{-2}$. These two
Kerr-Schild forms (2.43) and (2.44) prove the non-stationary
version of theorem 7 in the case of rotating Vaidya-Bonnor-Kerr
solution. The rotating Vaidya-Bonnor-Kerr metric will describe a
non-stationary spherically symmetric solution whose Weyl curvature
tensor is algebraically special in Petrov classification
possessing a geodesic, shear free, expanding and non-zero twist
null vector $\ell_a$. One can easily recover a rotating
Vaidya-Kerr metric from this Vaidya-Bonnor-Kerr solution by
setting the charge $e(u)=0$.

\vspace*{.15in}

\begin{center}
{\bf 3. Changing masses of non-stationary variable-charged black holes}
\end{center}
\setcounter{equation}{0}
\renewcommand{\theequation}{3.\arabic{equation}}

In this section, by solving Einstein-Maxwell field equations with the
{\sl variable}-charge $e(u,r)$, we develop the relativistic aspect
of Hawking radiation in non-stationary classical space-time metrics. The
calculation of Newman-Penrose (NP) spin coefficients is being
carried out through the technique developed by McIntosh and
Hickman [11] in (+,\,--,\,--,\,--) signature. In the formulation
of this relativistic aspect of Hawking radiation, we do not impose
any condition in the field equations except considering the charge
$e$ to be a function of the coordinates $u$ and $r$.
\vspace*
{0.2in}

{\sl (i) Variable-charged non-rotating Vaidya-Bonnor solution}\\\

We consider the non-rotating variable-charged Vaidya-Bonnor
solution with the assumption that the charge $e$ of the body is
a function of coordinate $u$ and $r$:
\begin{equation}
ds^2=\{1-{2M(u)\over r}+{e^2(u,r)\over
r^2}\}\,du^2+2du\,dr-r^2(d\theta^2
+{\rm sin}^2\theta\,d\phi^2).
\end{equation}
Initially when $e(u,r)=e(u)$, this metric provides the
non-rotating charged Vaidya-Bonnor solution [21]. Using the null tetrad
vectors one can calculate the spin coefficients, Ricci scalars and
Weyl scalars as follows:
\begin{eqnarray}
&&\kappa=\sigma=\nu=\lambda=\pi=\tau=\epsilon=0,\cr
&&\rho=-{1\over r},\;\ \beta=-\alpha={1\over {2\surd 2r}}\,{\rm
cot}\theta, \cr &&\mu=-{1\over 2\,r}\Big\{1-{2M(u)\over
r}+{e^2(u,r)\over r^2}\Big\},\;\;\ \cr &&\gamma={1\over
2\,r^2}\,\Big\{M(u)+e(u,r)\,e(u,r)_{,r}-{e^2(u,r)\over r^2}\Big\},\\
&&\phi_{00}=\phi_{01}=\phi_{10}=\phi_{20}=\phi_{02}=0,\cr
&&\phi_{11}={1\over 4\,r^2}\,\Big\{e(u,r)\,e(u,r)_{,r}\Big\}_{,r}
-{e(u,r)e(u,r)_{,r}\over r^3}+ {e^2(u,r)\over 2\,r^4},\cr
&&\phi_{22}={-1\over r^3}\Big\{r\,M(u)_{,u}-e(u,r)\,e(u,r)_{,u}\Big\} \cr
&&\Lambda=-{1\over{12\,r^2}}\,\Big\{e^2(u,r)_{,r}
+e(u,r)_{,r}\,e(u,r)_{,r\,r}\Big\},
\end{eqnarray}
\begin{eqnarray}
\psi_0&=&\psi_1=\psi_3=\psi_4=0,\cr \psi_2&=&{1\over
r^4}\,\Big[-r\,M(u)+e^2(u,r) - r\,e(u,r)\,e(u,r)_{,r},\cr
 &&+{1\over 6}\,r^2\,\Big\{e(u,r)\,e(u,r)_{,r}\Big\}_{,r}\Big],\;\
\end{eqnarray}
When the energy momentum tensor is of electromagnetic fields, then
the Ricci tensor $R_{ab}$ is proportional to the Maxwell stress
tensor [10] that is
\begin{equation}
\phi_{AB} = k\,\phi_A\,\overline \phi_B,\;\;\ k=8\pi G/c^2
\end{equation}
with $A,B$ = 0,1,2 and the NP Ricci scalar
\begin{equation}
\Lambda \equiv {1\over 24}\,R_{ab}\,g^{ab}=0.
\end{equation}
Hence, vanishing $\Lambda$ in (3.6) with (3.3) leads
\begin{equation}
e^2(u,r) = 2\,r\,m_1(u) + C(u)
\end{equation}
where $m_1(u)$ and $C(u)$ are real functions of $u$. Then the
Ricci scalar becomes
\begin{equation}
\phi_{11} ={C(u)\over 2\,r^4}.
\end{equation}
Thus, the Maxwell scalar $\phi_1={1\over 2}
F_{ab}(\ell^a\,n^b+\overline{m}^a\,m^b)$ takes the form, by
identifying the real function $C(u)$ = $e^2(u)$,
\begin{equation}
\phi_1={1\over\surd 2}\,e(u)\,r^{-2},
\end{equation}
showing that the Maxwell scalar $\phi_1$ does not change its form
by considering the charge $e$ to be a function of $u$ and $r$ in
Einstein-Maxwell field equations. Here, by using equation (3.7) in
(3.2) and (3.4), we have the resulting NP quantities
\begin{eqnarray}
&&\mu=-{1\over 2\,r}\Big[1-{2\over r}\{M(u) -
m_1(u)\}+{e^2(u)\over r^2}\Big],\\&&\gamma={1\over
2\,r^2}\,\Big[\{M(u) - m_1(u)\}-{e^2(u)\over r}\Big],\cr
&&\psi_2=-{1\over r^3}\,\Big[\{M(u) - m_1(u)\}-{e^2(u)\over
r}\Big].
\end{eqnarray}
and the metric (3.1) takes the form
\begin{equation}
ds^2=\Big[1-{2\over r}\Big\{M(u) - m_1(u)\Big\}+{e^2(u)\over
r^2}\Big]\,du^2 +2du\,dr-r^2(d\theta^2+{\rm
sin}^2\theta\,d\phi^2).
\end{equation}

This means that the mass $M(u)$ of non-rotating Vaidya-Bonnor
black hole (3.1) is lost a quantity $m_1(u)$ at the end of the
first electrical radiation. This loss of mass is agreeing with
Hawking's discovery that the radiating objects must lose its mass
[2]. On the lose of mass, Wald [22] has pointed that a black hole
will lose its mass at the rate as the energy is radiated. If one
considers the same process for second time taking $e$ in (3.12) to
be function of $u$ and $r$ with the mass $M(u) - m_1(u)$ in
Einstein-Maxwell field equations, then the mass may again be
decreased by another real function $m_2(u)$ (say); that is, after
the second time radiation the total mass might become $M(u) -
\{m_1(u) + m_2(u)\}$. This is due to the fact, that the Maxwell
scalar $\phi_1$ with condition (3.9) does not change its form
after considering the charge $e(u)$ to be function of $u$ and $r$
for the second time as $\Lambda$ calculated from the
Einstein-Maxwell field equations has to vanish for electromagnetic
fields with $e(u,r)$. Hence, if one repeats the same process for
$n$-times considering every time the charge $e(u)$ to be function
of $u$ and $r$, then one would expect the solution to change
gradually and the total mass becomes $M(u) - \{m_1(u) + m_2(u) +
m_3(u) + . . . + m_n(u)\}$ and therefore the metric (3.12) will
take the form:
\begin{equation}
ds^2=\Big[1-{2\over r}\,{\cal M}(u)+{e^2(u)\over r^2}\Big]\,du^2
+2du\,dr-r^2(d\theta^2+{\rm sin}^2\theta\,d\phi^2),
\end{equation}
where the mass of the black hole after the radiation of $n$-times would be
\begin{equation}
{\cal M}(u) = M(u) - \{m_1(u) + m_2(u) + m_3(u) + . . . +
m_n(u)\}.
\end{equation}
This suggests that for every electrical radiation, the original
mass $M(u)$ of the non-rotating black hole may lose some quantity.
Thus, it seems reasonable to expect that, taking Hawking's
radiation of black holes into account, such continuously  lose of
mass may lead to evaporate the original mass $M(u)$. In case the
black hole has evaporated down to the Planck mass, the mass $M(u)$
may not exactly equal to the continuously lost quantities $m_1(u)
+ m_2(u) + m_3(u) + . . .  + m_n(u)$. That is, according to the
second possibility of Hawking and Israel [3] quoted above, there
may left a small quantity of mass, say, Planck mass of about
$10^{-5}g$ with continuous electrical radiation. Otherwise, when
$M(u)$ = $m_1(u) + m_2(u) + m_3(u) + . . . + m_n(u)$ for a
complete evaporation of the mass, ${\cal M}(u)$ would be zero,
rather than, leaving behind a Planck-size mass black hole remnant.
At this stage the non-rotating black hole geometry would have the
electric charge $e$ only, but no mass, that the line element would
be of the form
\begin{equation}
ds^2=\Big(1+{e^2(u)\over
r^2}\Big)\,du^2+2du\,dr-r^2(d\theta^2+{\rm sin}^2\theta\,d\phi^2).
\end{equation}
Here, the metric describes the non-rotating charged solution
with imaginary roots $r_{\pm}=\pm\sqrt{-\{e^2(u)\}}$, indicating naked
singularity with zero mass. As the electrical radiation has to
continue, this remnant will remain only for an instant.

Here, again its electrical radiation will be continue to create
negative mass. [Steinmular, King and LoSota [6] refer a
spherically symmetric star, which radiates away all its mass as
`instantaneous' naked singularity that exists only for an instant
and then disappears. This `instantaneous' naked singularity is
also mentioned in [6]. The time taken between two consecutive
radiations is supposed to be very short that one may not
physically realize how quickly radiations take place. Thus it
seems natural to expect the existence of `instantaneous' naked
singularity with zero mass only for an instant before continuing
its next radiation to create negative mass naked singularity. It
may also be possible in the reasonable theory of black holes that,
as a black hole is invisible in nature, one may not know that in
the universe, a particular black hole has mass or not, but
electrical radiation may be detected on the black hole surface.
So, there may be some radiating black holes without masses in the
universe, where the gravity may depend only on the electric
charge, i.e., $\psi_2$=$e^2(u)/r^4$, not on the mass of the
black holes. Just after the exhaustion of the mass, if one
continues the remaining non-stationary solution (3.15) to radiate,
there may be a
formation of new mass $m^*_1(u)$ (say). If one repeats the
electrical radiation further, the new mass might increase
gradually and then, the metric (3.15) with the new mass would
become
\begin{equation}
ds^2=\Big(1+{2\over r}\,{\cal M}^*(u) + {e^2(u)\over
r^2}\Big)\,du^2+2du\,dr-r^2\Big(d\theta^2 +{\rm
sin}^2\theta\,d\phi^2\Big),
\end{equation}
where the new mass is given by
\begin{equation}
{\cal M}^*(u)=m^*_1(u) + m^*_2(u) + m^*_3(u)
 + m^*_4(u) + . . .
\end{equation}
Comparing the metrics (3.13) and (3.16) one could observe that the
classical space-time (3.16) will describe a non-rotating spherical
symmetric star with a negative mass ${\cal M}^*(u)$. Such objects
with negative masses are referred as naked singularities [1, 6, 7].
The metric (3.16) may be regarded to describe the incorporation of
the third possibility of Hawking and Israel [3] in the case of
non-rotating singularity. Here it is noted that the creation of
negative mass naked singularity is mainly based on the continuous
electrical radiation of the variable charge $e(u,r)$ in the energy
momentum tensor of Einstein-Maxwell equations. This also indicates
the incorporation of Boulware's suggestion [5] that `the
stress-energy tensor may be used to calculate the change in the
metric due to the radiation'. This new mass ${\cal M}^*(u)$ would
never decrease, rather might increase gradually as the radiation
continues forever. and the unaffected Maxwell scalar $\phi_1$ is
given in (3.9). The metric (3.16) admits the energy momentum
tensor with the mass ${\cal M}^{*}$ showing effect of radiation.
Thus, one has seen the changes in the mass of the
non-rotating charged black hole after every radiation. Hence, it
follows the theorem 1 cited above in the case of non-rotating {\sl
variable-charged} black hole.

\vspace* {0.2in}

{\sl (ii) Variable-charged rotating Vaidya-Bonnor Black hole} \\\

Here, we shall incorporate the Hawking radiation, how the
variable-charged rotating black hole affect in the classical
space-time metric when the electric charge $e$ is taken as a
function of $u$ and $r$ in the Einstein-Maxwell field equations.
The line element with $e(u,r)$ is
\begin{eqnarray}
ds^2&=&[1-R^{-2}\{2rM(u)-e^2(u,r)\}]\,du^2+2du\,dr \cr
&+&2aR^{-2}\{2rM(u)-e^2(u,r)\}\,{\rm sin}^2
\theta\,du\,d\phi-2a\,{\rm sin}^2\theta\,dr\,d\phi \cr
&-&R^2d\theta^2-\{(r^2+a^2)^2 -\Delta^*a^2\,{\rm
sin}^2\theta\}\,R^{-2}{\rm sin}^2\theta\,d\phi^2,
\end{eqnarray}
where $\Delta^*=r^2-2rM(u)+a^2+e^2(u,r)$. This metric will also
reduce to rotating Vaidya-Bonnor solution when $e(u,r)=e(u)$
initially, then the Einstein-Maxwell field equations for the
metric (3.18) can be solved to obtain the following change NP
quantities
\begin{eqnarray}
&&\gamma={1\over{2\overline R\,R^2}}\,\left[\{r-M(u)+
e(u,r)\,e(u,r)_{,r}\}\overline R-\Delta^*\right],\cr
&&\nu={1\over{\surd 2\overline R\,R^2}}\,\Big[i\,a\,{\rm
sin}\theta\,\Big\{r\,M(u)_{,u} -e(u,r)\,e(u,r)_{,u}\Big\}\Big].
\nonumber
\end{eqnarray}
The Weyl scalars are
\begin{eqnarray}
\psi_0&=&\psi_1=0,\cr \psi_2 &=&{1\over\overline R\,\overline
R\,R^2}\Big[\Big\{-RM(u)+e^2(u,r)\Big\}-\overline
R\,e(u,r)\,e(u,r)_{,r}\cr && +{1\over 6}\,\overline R\,\overline
R\,\Big\{e^2(u,r)_{,r}+e(u,r)\,e(u,r)_{,r\,r}\Big\}\Big],\\ \psi_3
&=&-{1\over{2\surd 2\overline R\,\overline R\,R^2}}\,
\Big[4\,i\,a\,{\rm
sin}\theta\,\Big\{rM(u)_{,u}-e(u,r)\,e(u,r)_{,u}\Big\} \cr &&+
i\,a\,{\overline R}\,{\rm sin}\theta\Big\{M(u)
-e(u,r)\,e(u,r)_{,r}\Big\}_{,u}\Big],\cr \psi_4 &=&{a^2\,{\rm
sin}^2\theta\over{2\overline R\,\overline
R\,R^2}}\,\Big\{rM(u)_{,u}-e(u,r)\,e(u,r)_{,u}\Big\}_{,u} \cr
&&-{a^2\,r\,{\rm sin}^2\theta\over{\overline R\,\overline
R\,R^2\,R^2}}\Big[r\,M(u)_{,u}-e(u,r)\,e(u,r)_{,u}\Big].
\end{eqnarray}
and the Ricci scalars:
\begin{eqnarray}
\phi_{00}&=&\phi_{01}=\phi_{10}=\phi_{20}=\phi_{02}=0,\cr
\phi_{11}&=&{1\over{4\,R^2\,R^2}}\,\Big[ 2\,e^2(u,r)
-4re(u,r)\,e(u,r)_{,r}+R^2\{e(u,r)\,e(u,r)_{,r}\}_{,r}\Big],\\
\phi_{12}&=&{i\,
 a\,{\rm sin}\theta\over{2\surd 2\,R^2\,R^2}}
\,\Big[RM(u)_{,u}-2e(u,r)e(u,r)_{,u}+\overline
R\{e(u,r)e(u,r)_{,u}\}_{,r} \Big], \cr
\phi_{22}&=&{-1\over{R^2\,R^2}}\,\Big[r\Big\{r\,M(u)_{,u}-e(u,r)\,e(u,r)_{,u}\Big\}\Big],\cr
&&-{a^2\,{\rm
sin}^2\theta\over{R^2\,R^2}}\,\Big[r\,M(u)_{,u}-\Big\{e(u,r)\,e(u,r)_{,u}\Big\}\Big]_{,u},\cr
\Lambda
&=&{-1\over{12\,R^2}}\,\Big\{e^2(u,r)_{,r}+e(u,r)\,e(u,r)_{,r\,r}.
\end{eqnarray}
For electromagnetic field the Ricci scalar $\Lambda$ must vanish.
Thus, the vanishing $\Lambda$ of (3.22) implies that
\begin{equation}
e^2(u,r) = 2\,rm_1(u)\ + C(u)
\end{equation}
where $m_1(u)$ and $C(u)$ are real functions of $u$. Then, using
this result in equation (3.21) we obtain the Ricci scalar
\begin{equation}
\phi_{11}={C(u)\over 2\,R^2\,R^2}.
\end{equation}
Accordingly, the Maxwell scalar will become, after identifying the
function $C(u)=e^2(u)$,
\begin{equation}
\phi_1={e(u)\over\surd 2\,\overline R\,\overline R}.
\end{equation}
This shows that the Maxwell scalar $\phi_1$ is unaffected by
considering the charge $e(u)$ to be a function $u, r$ in the field
equations. Hence, from the Einstein-Maxwell equations we obtain
the changed NP quantities
\begin{eqnarray}
\mu&=&-{1\over{2\overline
R\,R^2}}\Big[r^2-2\,r\{M(u)-m_1(u)\}+a^2+e^2(u)\Big],\\
\gamma&=&{1\over{2\overline
R\,R^2}}\,\Big[\Big\{r-(M(u)-m_1(u))\Big\}\overline R \cr
&&-\Big\{r^2-2r\Big(M(u)-m_1(u)\Big)+a^2+e^2(u)\Big\}\Big],\cr
\psi_2&=&{1\over{\overline R\,\overline
R\,R^2}}\Big[-\{M(u)-m_1(u)\}R + e^2(u)\Big],\\
\psi_3&=&{-1\over{2\surd 2\overline R\,\overline
R\,R^2}}\,\Big[i\,a\,{\rm sin}\theta\,\Big\{(4r+\overline
R)\Big(M(u)-m_1(u)\Big)_{,u}\Big\} \cr &&-4\,i\,a\,{\rm
sin}\theta\,\Big\{e(u)\,e(u)_{,u}\Big\}\Big],\cr
\psi_4&=&{a^2\,{\rm sin}^2\theta\over{2\overline R\,\overline
R\,R^2}}\,\Big[r\,\Big\{M(u)-m_1(u)\Big\}_{,u}-e(u)e(u)_{,u}\Big]_{,u},\cr
&&-{a^2\,r\,{\rm sin}^2\theta\over{\overline R\,\overline
R\,R^2\,R^2}}\,\Big[r\,\Big\{M(u)-m_1(u)\Big\}_{,u}-e(u)e(u)_{,u}\Big],\\
\phi_{11}&=&{e^2(u)\over 2\,R^2\,R^2}. \nonumber
\end{eqnarray}
Thus, the solution (3.18) with a new function $m_1(u)$ takes the
following form
\begin{eqnarray}
ds^2&=&\Big[1-R^{-2}\Big\{2r\Big(M(u)-m_1(u)\Big)-e^2(u)\Big\}\Big]\,du^2+2du\,dr
\cr &+&2aR^{-2}\Big\{2r\Big(M(u)-m_1(u)\Big)-e^2(u)\Big\}\,{\rm
sin}^2 \theta\,du\,d\phi-2a\,{\rm sin}^2\theta\,dr\,d\phi \cr
&-&R^2d\theta^2-\{(r^2+a^2)^2 -\Delta^*a^2\,{\rm
sin}^2\theta\}\,R^{-2}{\rm sin}^2\theta\,d\phi^2,
\end{eqnarray}
where $\Delta^*=r^2-2r\{M(u)-m_1(u)\}+a^2+e^2(u)$. This
introduction of mass $m_1(u)$ in the metric (3.29) suggests that
the first electrical radiation of rotating black hole may reduce
the original gravitational mass $M(u)$ by a quantity $m_1(u)$. If
one considers another radiation by taking $e(u)$ in (3.29) to be a
function of $u$ and $r$ with the mass $M(u)-m_1(u)$, then the
Einstein-Maxwell field equations yield to reduce this mass by
another real function $m_2(u)$; i.e., after the second radiation,
the mass may become $M(u)-\{m_1(u)+m_2(u)\}$. Here again, the
Maxwell scalar $\phi_1$ remains the same form after the second
radiation also. Thus, if one considers the $n$-time radiations
taking every time the charge $e(u)$ to be function of $u, r$, the
Maxwell scalar $\phi_1$ will remain unaffected, but the metrics
will be of the following form:
\begin{eqnarray}
ds^2&=&[1-R^{-2}\{2r{\cal M}(u)-e^2(u)\}]\,du^2+2du\,dr \cr
&&+2aR^{-2}\{2r{\cal M}(u)-e^2(u)\}\,{\rm sin}^2
\theta\,du\,d\phi-2a\,{\rm sin}^2\theta\,dr\,d\phi \cr
&&-R^2d\theta^2-\{(r^2+a^2)^2 -\Delta^*a^2\,{\rm
sin}^2\theta\}\,R^{-2}{\rm sin}^2\theta\,d\phi^2,
\end{eqnarray}
where the total mass of the black hole, after the $n$-time
radiations will take the form
\begin{equation}
{\cal M}(u)=M(u)-\{m_1(u) + m_2(u) + m_3(u) + m_4(u) + . . .+
m_n(u)\}.
\end{equation}
Taking Hawking's radiation of black holes, one might expect that
the total mass of  black hole may be radiated away just leaving
${\cal M}(u)$ equivalent to Planck mass of about $10^{-5}g$, that
is, $M(u)$ may not be exactly equal to $m_1(u) + m_2(u) + m_3(u) +
m_4(u) + . . .+ m_n(u)$, but has a difference of about Planck-size
mass, as in the case of non-rotating black hole given in (i).
Otherwise, the
total mass of black hole may be evaporated completely after
continuous radiation when ${\cal M}(u) = 0$, that is, $M(u) =
m_1(u) + m_2(u) + m_3(u) + m_4(u) + . . .+ m_n(u)$. Here one may
regard that the rotating variable-charged black hole might be
radiated completely away all its mass just leaving the electrical
charge $e(u)$ only. One could observe this situation in the form
of classical space-time metric as
\begin{eqnarray}
ds^2&=&(1+e^2(u)\,R^{-2})\,du^2+2du\,dr \cr &-&2ae^2\,R^{-2}\,{\rm
sin}^2 \theta\,du\,d\phi-2a\,{\rm sin}^2\theta\,dr\,d\phi \cr
&-&R^2d\theta^2-\{(r^2+a^2)^2 -\Delta^*a^2\,{\rm
sin}^2\theta\}\,R^{-2}{\rm sin}^2\theta\,d\phi^2,
\end{eqnarray}
with the charge $e(u)$, but no mass, where
$\Delta^*=r^2+a^2+e^2(u)$. The metric (3.32) will describe a
rotating `instantaneous' naked singularity with zero mass. At this
stage, the Weyl scalar $\psi_2$, $\psi_3$ and $\psi_4$ takes the
form
\begin{eqnarray}
&&\psi_2={e^2(u)\over{\overline R\,\overline R\,R^2}}, \\
&&\psi_3={\surd 2\,a\,i\,{\rm sin}\theta\over{\overline R
\,\overline R\,R^2}}\,\{e(u)e(u)_{,u}\},\\ &&\psi_4={a^2\,{\rm
sin}^2\theta\over{2\,\overline R\,\overline
R\,R^2\,R^2}}\,[2\,r\,e(u)e(u)_{,u}-R^2\,\{e(u)e(u)_{,u}\}_{,u}].
\end{eqnarray}
showing the gravity on the surface of the remaining solution
depending only on the electric charge $e(u)$; however, the Maxwell
scalar $\phi_1$ remains the same as in (3.25). For future use, we
mention the changed NP spin coefficients
\begin{eqnarray}
&&\mu=-{1\over{2\overline R\,R^2}}\{r^2+a^2+e^2(u)\},\cr
&&\gamma={1\over{2\overline R\,R^2}}\,[r\overline
R-\{r^2+a^2+e^2(u)\}],\\
&&\phi_{11}={e^2(u)\over{2\,R^2\,R^2}},\\
&&\phi_{12}={-i\,
 a\,{\rm sin}\theta\over{\surd 2\,R^2\,R^2}}
\,\Big\{e(u)e(u)_{,u}\Big\}, \cr
&&\phi_{22}={1\over{R^2\,R^2}}\,\Big\{r\,e(u)\,e(u)_{,u}\Big\},
+{a^2\,{\rm
sin}^2\theta\over{R^2\,R^2}}\,\Big\{e(u)\,e(u)_{,u}\Big\}_{,u}.
\end{eqnarray}
As the electrical radiation has to continue, the remaining remnant
will remain only for an instant. Hence we refer to the solution
(3.32) as an 'instantaneous' naked singularity with zero mass.
It suggests that there may be rotating black holes in the universe
whose masses are completely radiated; their gravity depend only on
the electric charge of the body and their metrics look like the
one given in the equation (3.32). It appears that the idea of this
evaporation of masses of radiating black holes may be agreed with
that of Hawking's evaporation of black holes. Unruh [23] has
examined various aspects of black hole evaporation based on
Schwarzschild metric. It is worthwhile to study the nature of such
black holes (3.32). This might give a different nature, which one
has not yet come across so far in the reasonable theory of black
holes. Here, immediately after the exhaustion of the Vaidya-Bonnor
mass, one may consider again the charge $e(u)$ to be function of
$u$ and $r$ for next radiation in (3.32), so that one must get
from the Einstein's field equations the scalar $\Lambda$ as given
in equation (3.22). Then the vanishing of this $\Lambda$ for
electromagnetic field, there will be creation of a new mass (say
$m^*_1(u)$) in the remaining space-time geometry. If this
radiation process continues forever, the new mass will increase
gradually as
\begin{equation}
{\cal M}^*(u)=m^*_1(u) + m^*_2(u) + m^*_3 (u)+ m^*_4(u) + . . .
.\,\,
\end{equation}
However, it appears that this new mass would never decrease. Then
the space-time geometry takes the form
\begin{eqnarray}
ds^2&=&[1+R^{-2}\{2r{\cal M}^*(u)+e^2(u)\}]\,du^2+2du\,dr \cr
&-&2aR^{-2}\{2r{\cal M}^*(u)+e^2(u)\}\,{\rm sin}^2
\theta\,du\,d\phi-2a\,{\rm sin}^2\theta\,dr\,d\phi \cr
&-&R^2d\theta^2-\{(r^2+a^2)^2 -\Delta^*a^2\,{\rm
sin}^2\theta\}\,R^{-2}{\rm sin}^2\theta\,d\phi^2,
\end{eqnarray}
where $\Delta^*=r^2+2r{\cal M}^*(u)+a^2+e^2(u)$. The affected Weyl
scalars and other NP coefficients are calculated from the
Einstein-Maxwell field equations as
\begin{eqnarray}
\psi_2&=&{1\over{\overline R\,\overline R\,R^2}}\,\Big\{{\cal
M}^*(u)R+e^2(u)\Big\}.\\
\psi_3&=&{a\,i\,{\rm sin}\theta\over{2\surd 2\overline
R\,\overline R\,R^2}}\,\Big\{(4r+\overline R){\cal
M}^*(u)_{,u}+4\,e(u)\,e(u)_{,u}\Big\},\\ \psi_4&=&-{a^2\,{\rm
sin}^2\theta\over{2\overline R\,\overline R\,R^2}}\,\Big\{r{\cal
M}^*(u)+e(u)\,e(u)_{,u}\Big\}_{,u},\cr &&+{r\,a^2\,{\rm
sin}^2\theta\over{\overline R\,\overline
R\,R^2\,R^2}}\,\Big\{r{\cal M}^*(u)+e(u)e(u)_{,u}\Big\},\\
\mu&=&-{1\over{2\overline R\,R^2}}\Big\{r^2
+2r{\cal M}^*(u)+a^2+e^2(u)\Big\},\\
\gamma&=&{1\over{2\overline R\,R^2}}\,\Big[\Big\{\Big(r+{\cal
M}^*(u)\Big)\overline R-\Big\{r^2+2r{\cal
M}^*(u)+a^2+e^2(u)\Big\}\Big], \nonumber
\end{eqnarray}
with $\phi_1$ remained the same as in (3.25). The metric (3.40) may be regarded to
describe a rotating negative mass naked singularity. We have
presented the possible changes in the mass of the rotating charged
black hole without affecting the Maxwell scalar $\phi_1$ and
accordingly, metrics are cited for future use. Thus, this
completes the proof of the theorem 4 for the rotating part of charged
black hole.

\vspace*{0.2in}

{\it (iii) Variable-charged rotating Vaidya-Bonnor-de Sitter Black hole}\\\

In this section we shall consider the variable-charged black hole
embedded into de Sitter space. By solving Einstein-Maxwell
field equations with the variable-charge $e(u,r)$
of rotating Vaidya-Bonnor black holes embedded in de Sitter
space, we develop the relativistic aspect of Hawking radiation in
classical spacetime metrics. We consider the charge $e$ to be
function of coordinates $u$ and $r$ and the decomposition of the
Ricci scalar $\Lambda\equiv(1/24)\,R_{ab}\,g^{ab}$ into two parts,
without loss of generality, as follows
\begin{equation}
\Lambda=\Lambda^{(\rm C)}+\Lambda^{(\rm E)},
\end{equation}
where $\Lambda^{(\rm C)}$ is the {\it non-zero} cosmological Ricci
scalar and $\Lambda^{(\rm E)}$ is the {\it zero} Ricci scalar of
electromagnetic field for the rotating as well as non-rotating
black holes. This decomposition of Ricci scalar $\Lambda$ is
possible because the cosmological object and the electromagnetic
field are two different matter fields of different physical
nature, though they are supposed to exist on the same spacetime
coordinates here. For our purpose of the paper, this type of
decomposition of Ricci scalars $\Lambda$ will serve well in the
study of Hawking's radiation of black holes embedded into the de
Sitter cosmological space. The line element of rotating
Vaidya-Bonnor-de Sitter black hole with variable charge $e(u,r)$
is
\begin{eqnarray}
ds^2&=&\Big[1-R^{-2}\Big\{2rM(u)-e^2(u,r)+{\Lambda^*r^4\over
3}\Big\}\Big]\,du^2+2du\,dr \cr
&&+2aR^{-2}\Big\{2rM(u)-e^2(u,r)+{\Lambda^*r^4\over 3}\Big\}\,{\rm
sin}^2 \theta\,du\,d\phi-2a\,{\rm sin}^2\theta\,dr\,d\phi \cr
&&-R^2d\theta^2-\Big\{(r^2+a^2)^2 -\Delta^*a^2\,{\rm
sin}^2\theta\Big\}\, R^{-2}{\rm sin}^2\theta\,d\phi^2,
\end{eqnarray}
where $\Delta^*=r^2-2rM(u)+a^2+e^2(u,r)-\Lambda^*r^4/3$. This
metric will recover the rotating Vaidya-Bonnor-de Sitter solution
given in (2.3) when $e(u,r)=e(u)$ initially. Then the Einstein-Maxwell
field equations for the metric (3.46) can be solved. So we obtain
the changed NP quantities
\begin{eqnarray}
\mu&=&-{1\over{2\overline R\,R^2}}\Big\{r^2-2rM(u)+a^2+e^2(u,r)
-{\Lambda^*r^4\over 3}\Big\},\\ \gamma&=&{1\over{2\overline
R\,R^2}}\,\Big[\Big\{r-M(u)+e(u,r)\,e(u,r)_{,r}-{\Lambda^*r^4\over
3} \Big\}\overline R \cr
&&-\Big\{r^2-2rM(u)+a^2+e^2(u,r)-{2\Lambda^*r^4\over
3}\Big\}\Big],\\\psi_2&=&{1\over{\overline R\,\overline
R\,R^2}}\,\Big\{-RM(u)+ e^2(u,r)-e(u,r)e(u,r)_{,r}\,{\overline
R}+{\Lambda^*r^2\over 3} a^2\,{\rm cos}^2\theta\Big\} \cr
&&+{1\over{6\,R^2}}\Big\{e(u,r)\,e(u,r)_{,r}\Big\}_{,r},\\
\phi_{11}&=&{1\over
{2\,R^2\,R^2}}\,\Big\{e^2(u,r)-2r\,e(u,r)e(u,r)_{,r}
-{\Lambda^*r^2}\,a^2\,{\rm cos}^2\theta\Big\} \cr && + {1\over
4R^2}\,\Big\{e^2(u,r)_{,r}+e(u,r)\,e(u,r)_{,r\,r}\Big\},\\
\Lambda &=&{\Lambda^*r^2\over 6\,R^2} -{1\over
12\,R^2}\,\Big\{e^2(u,r)_{,r}+e(u,r)\,e(u,r)_{,r\,r}\Big\}.
\end{eqnarray}
where $\psi_3$ and $\psi_4$ are same as (3.20) and $\phi_{12}$,
$\phi_{22}$ in (3.21)of case $ 3(ii)$. Thus, we have seen that in
each expression of $\phi_{11}$ and $\psi_2$ there is the cosmological
$\Lambda^*$ term coupling with the rotation parameter $a$. Hence,
without loss of generality, it will be convenient here to have a
decomposition of $\phi_{11}$ into two parts - one for the cosmological
Ricci scalar $\phi_{11}^{(\rm C)}$ and the other for the electromagnetic
field $\phi_{11}^{(\rm E)}$ as in the case of $\Lambda$ in (3.45), such
that
\begin{eqnarray}
\phi_{11}^{(\rm C)}&=&-{1\over {2\,R^2\,R^2}}\,
{\Lambda^*r^2}\,a^2\,{\rm cos}^2\theta,\\ \phi_{11}^{(\rm
E)}&=&{1\over
{2\,R^2\,R^2}}\,\Big\{e^2(u,r)-2r\,e(u,r)e(u,r)_{,r}\Big\},\cr &&+
{1\over 4R^2}\,\Big\{e(u,r)\,e(u,r)_{,r}\Big\}_{,r}.
\end{eqnarray}
Similarly, we also have the decomposition of $\Lambda$ as
\begin{eqnarray}
&&\Lambda^{(\rm C)} ={\Lambda^*r^2\over 6\,R^2},\\ &&
\Lambda^{(\rm E)} =-{1\over
12\,R^2}\,\Big\{e^2(u,r)_{,r}+e(u,r)\,e(u,r)_{,r\,r}\Big\}.
\end{eqnarray}
Now, the scalar $\Lambda^{(\rm E)}$ for electromagnetic field must
vanish for this rotating metric. Thus, the vanishing
$\Lambda^{(\rm E)}$ of the equation (3.55) yields that
\begin{equation}
e^2(u,r) = 2\,rm_1(u) + C(u)
\end{equation}
where $m_1(u)$ and $C(u)$ are real functions. Then, substituting
this result in equation (3.53) we obtain the Ricci scalar for
electromagnetic field
\begin{equation}
\phi_{11}^{(\rm E)}={C(u)\over 2\,R^2\,R^2}.
\end{equation}
However, the cosmological Ricci scalar $\phi_{11}^{(\rm C)}$
remains the same form as in (3.52). Accordingly, the Maxwell
scalar takes, after identifying the function $C(u)\equiv e^2(u)$,
\begin{equation}
\phi_1={e(u)\over\surd 2\,\overline R\,\overline R}.
\end{equation}
Hence, the affected NP quantities after substitution of $e^2(u,r)$ (3.56) are
\begin{eqnarray}
\mu&=&-{1\over{2\overline
R\,R^2}}\Big\{r^2-2r\Big(M(u)-m_1(u)\Big)+a^2
+e^2(u)-{\Lambda^*r^4\over 3}\Big\},\\
\gamma&=&{1\over{2\overline
R\,R^2}}\,\Big[\Big\{r-\Big(M(u)-m_1(u)\Big)-{\Lambda^*r^4\over 3}
\Big\}\overline R \cr &&-\Big\{r^2-2r\Big(M(u)-m_1(u)\Big)
+a^2+e^2(u)-{2\Lambda^*r^4\over 3}\Big\}\Big],\cr
\psi_2&=&{1\over{\overline R\,\overline
R\,R^2}}\,\Big\{-\Big(M(u)-m_1(u)\Big)R+
e^2(u)+{\Lambda^*r^2\over3}\,a^2\,{\rm cos}^2\theta\Big\},\\
\phi_{11}&=&{1\over {2\,R^2\,R^2}}\,\Big\{e^2(u)
-{\Lambda^*r^2}\,a^2\,{\rm cos}^2\theta\Big\},\\
\Lambda&=&\Lambda^{(\rm C)} ={\Lambda^*r^2\over 6\,R^2}.
\end{eqnarray}
We have seen the changes in $\mu$, $\gamma$, $\psi_2$, $\psi_3$
and $\psi_4$ but changes in $\psi_3$ and $\psi_4$ are same as
(3.28) of case 3(ii) and there is no change in $\phi^{(\rm
C)}_{11}$ and $\Lambda^{(\rm C)}$. Thus, the rotating solution
(3.46) with a new real function $m_1(u)$ after the first radiation
becomes
\begin{eqnarray}
ds^2&=&\Big[1-R^{-2}\Big\{2r\Big(M(u)-m_1(u)\Big)-e^2(u)+{\Lambda^*r^4\over
3}\Big\}\Big]\,du^2 +2du\,dr \cr
&&+2aR^{-2}\Big\{2r\Big(M(u)-m_1(u)\Big)-e^2(u)+{\Lambda^*r^4\over
3}\Big\}\,{\rm sin}^2 \theta\,du\,d\phi-2a\,{\rm
sin}^2\theta\,dr\,d\phi \cr &&-R^2d\theta^2-\Big\{(r^2+a^2)^2
-\Delta^*a^2\,{\rm sin}^2\theta\Big\}\, R^{-2}{\rm
sin}^2\theta\,d\phi^2,
\end{eqnarray}
where
$\Delta^*=r^2-2r\Big\{M(u)-m_1(u)\Big\}+a^2+e^2(u)-\Lambda^*r^4/3$.
This suggests that the first electrical radiation of rotating
black hole leads to a reduction of the gravitational mass $M(u)$
by a quantity $m_1(u)$ with the unaffected Maxwell scalar $\phi_1$
and the constant $\Lambda^{(\rm C)}$. If we consider another
radiation by taking $e(u)$ in (3.63) to be a function of $u$ and
$r$ with the mass $M(u)-m_1(u)$ and the decomposition (3.45), then
the Einstein-Maxwell field equations yield to reduce this mass by
another quantity $m_2(u)$ (say); i.e., after the second radiation,
the mass will become $M(u)-\{m_1(u)+m_2(u)\}$. Here again, the
Maxwell scalar $\phi_1$ and the constant $\Lambda^{(\rm C)}$
remain unaffected after the second radiation also. Thus,
if we consider $n$ radiations everytime taking the charge $e$ to
be function of $u$ and $r$ with the decomposition of $\Lambda$,
the Maxwell scalar $\phi_1$ will be the same, but the metric will
take the form:
\begin{eqnarray}
ds^2&=&\Big[1-R^{-2}\Big\{2r{\cal M}(u)-e^2(u)+{\Lambda^*r^4\over
3}\Big\}\Big]\,du^2+2du\,dr \cr &&+2aR^{-2}\Big\{2r{\cal
M}(u)-e^2(u)+{\Lambda^*r^4\over 3}\Big\}\,{\rm sin}^2
\theta\,du\,d\phi-2a\,{\rm sin}^2\theta\,dr\,d\phi \cr
&&-R^2d\theta^2-\Big\{(r^2+a^2)^2 -\Delta^*a^2\,{\rm
sin}^2\theta\Big\}\,R^{-2}{\rm sin}^2\theta\,d\phi^2,
\end{eqnarray}
where the total mass of the black hole, after the $n$ radiations
will be of the form
\begin{equation}
{\cal M}(u)=M(u)-\{m_1(u) + m_2(u) + m_3(u) + m_4(u) + . . .+
m_n(u)\}.
\end{equation}
Taking Hawking's radiation of black holes, we can expect that the
total mass of  black hole will be radiated away just leaving
${\cal M}(u)$ equivalent to Planck mass of about $10^{-5}$ g, that
is, $M(u)$ may not be exactly equal to $m_1(u) + m_2(u) + m_3(u) +
m_4(u) + . . .+ m_n(u)$, but has a difference of about Planck-size
mass, as in the case of non-rotating black hole. Otherwise, the
total mass of black hole will be evaporated completely after
continuous radiation when ${\cal M}(u) = 0$, that is, $M(u) =
m_1(u) + m_2(u) + m_3(u) + m_4(u) + . . .+ m_n(u)$. Here the
rotating variable-charged black hole might completely radiate away
its mass just leaving the electrical charge $e(u)$ and the
cosmological constant $\Lambda^*$. We find this situation in the
form of classical space-time metric as
\begin{eqnarray}
ds^2&=&\Big\{1+\Big(e^2(u)-{\Lambda^*r^4\over
3}\Big)\,R^{-2}\Big\}\,du^2+2du\,dr \cr
&&-2a\,R^{-2}\,\Big(e^2(u)-{\Lambda^*r^4\over 3}\Big)\,{\rm sin}^2
\theta\,du\,d\phi-2a\,{\rm sin}^2\theta\,dr\,d\phi \cr
&&-R^2d\theta^2-\Big\{(r^2+a^2)^2 -\Delta^*a^2\,{\rm
sin}^2\theta\Big\}\,R^{-2}{\rm sin}^2\theta\,d\phi^2,
\end{eqnarray}
with the charge $e(u)$ and the cosmological constant $\Lambda^*$,
but no mass, where $\Delta^*=r^2+a^2+e^2(u)-\Lambda^*r^4/3$. The
metric (3.66) will describe a rotating `instantaneous' charged
cosmological black holes. At this stage, the Weyl scalar $\psi_2$
takes the form
\begin{eqnarray}
\psi_2&=&{1\over{\overline R\,\overline R\,R^2}}\Big\{
e^2(u)+{\Lambda^*r^2\over3}\,a^2\,{\rm cos}^2\theta\Big\},
\end{eqnarray}
where as changed in $\psi_3$ and $\psi_4$ are remain same as in
(3.34) and (3.35). At this stage we shall mention
some important parameters of the charged cosmological black
hole. That is, the surface gravity of the horizon at $r=r_{+\,+}$
is
\begin{eqnarray}
{\cal
K}={-\Big[{1\over{r\,R^2}}\,\Big\{r\,\Big(r-{\Lambda^*r^3\over 6
}\Big)+{e^2(u)\over 2}\Big\}\Big]}_{r=r_{+\,+}},
\end{eqnarray}
with the Hawking's temperature ${\cal T}_{\rm H}={\cal K
}/{2\,\pi}$. The entropy and angular velocity of the black hole at
the horizon are found as follows
\begin{eqnarray}
{\cal S}=\pi\,(r^2+a^2)_{r=r_{+\,+}},\;\;\; {\rm and}\;\;\;
\Omega_{\rm H} ={-a\,\{e^2(u)-{\Lambda^*\,r^4/
3}\}\over(r^2+a^2)^2}\Big|_{r=r_{+\,+}}.
\end{eqnarray}
Here, the value of $r_{\pm}$ may be obtained from appendix (A15).
At this instant, the gravity on the surface of the remaining
solution depending on the electric charge $e(u)$ and the
cosmological constant $\Lambda^*$ coupling with the rotational
parameter $a$ can be seen in (3.67); however, the Maxwell scalar
$\phi_1$ and the Ricci scalar $\Lambda^{(\rm C)}$ remain the same
as in (3.58) and (3.54) respectively. The formation of charged
cosmological black hole (3.66) leads the proof of one part of the
theorem 3 cited above for the case embedded into de
Sitter space. Here the idea of this complete evaporation of mass
of radiating black hole embedded into the de Sitter space is in
agreement with that of Hawking's evaporation of black holes. It is
worth studying the nature of such rotating black holes (3.66).
This might give a different physical nature, which one has not
been seen in common theory of black holes embedded into the de
Sitter space. Here, we again consider the charge $e(u)$ to be
function of $u$ and $r$ for the next radiation in (3.66), so that
we get, from the Einstein's field equations, the scalar
$\Lambda^{(\rm E)}$ as given in equation (3.55) and the same
scalar $\Lambda^{(\rm C)}$ as in (3.54). Then the vanishing of
this $\Lambda^{(\rm E)}$ for electromagnetic field will lead to
create a new mass (say $m^*_1(u)$) in the remaining space-time
geometry (3.66). For the second radiation, we again consider the
charge $e(u)$ to be function of $u$ and $r$ in the field equations
with the mass $m^*_1(u)$. Then the vanishing of $\Lambda^{(\rm
E)}$ will lead to increase the new mass by another quantity
$m^*_2(u)$ (say) i.e., after the second radiation of (3.66) the
new mass will be $m^*_1(u) +m^*_2(u)$. If this radiation process
continues further for a long time, the new mass will increase
gradually as
\begin{equation}
{\cal M}^*(u)=m^*_1(u) + m^*_2(u) + m^*_3(u) + m^*_4(u) + . . .
.\,\,
\end{equation}
Then, the spacetime metric will take the form
\begin{eqnarray}
ds^2&=&\Big[1+R^{-2}\Big\{2r{\cal
M}^*(u)+e^2(u)-{\Lambda^*r^4\over 3}\Big\}\Big]\,du^2+2du\,dr \cr
&&-2aR^{-2}\Big\{2r{\cal M}^*(u)+e^2(u)-{\Lambda^*r^4\over
3}\Big\}\,{\rm sin}^2 \theta\,du\,d\phi-2a\,{\rm
sin}^2\theta\,dr\,d\phi \cr &&-R^2d\theta^2-\Big\{(r^2+a^2)^2
-\Delta^*a^2\,{\rm sin}^2\theta\Big\}\, R^{-2}{\rm
sin}^2\theta\,d\phi^2,
\end{eqnarray}
where $\Delta^*=r^2+2r{\cal M}^*(u)+a^2+e^2(u)-{\Lambda^*r^4/3}$.
The changed NP quantities $\psi_2$, $\psi_3$, $\psi_4$ and $\mu$
are as follows
\begin{eqnarray}
&&\psi_2={1\over{\overline R\,\overline R\,R^2}}\,\Big\{{\cal
M}^*(u)R+e^2(u)+{\Lambda^*r^2\over3}\,a^2\,{\rm cos}^2\theta\},\\
&&\mu=-{1\over{2\overline R\,R^2}}\Big\{r^2+2r{\cal
M}^*(u)+a^2+e^2(u)-{\Lambda^*r^4\over 3}\Big\},
\end{eqnarray}
with $\psi_3$ and $\psi_4$, given as in (3.43), and $\phi_1$
and $\Lambda^{(\rm C)}$ remain same as in (3.58) and(3.54).\\
Comparing the metrics (3.64) and (3.71), we find that the classical
spacetime (3.71) describes a rotating negative mass naked singularity
embedded into the de Sitter cosmological space. Thus, from the above
it follows the proof of theorem 5 in the case of de Sitter space.
We have also shown the possible changes in the mass of the rotating charged
Vaidya-Bonnor-de Sitter black hole without affecting the Maxwell
scalar $\phi_1$ as well as the cosmological constant $\Lambda^*$
and accordingly, metrics are cited for future use. Thus, this
completes the proof of the theorem 1 based on the rotating charged
de Sitter black hole. Also since there is no effect on the
cosmological constant $\Lambda^*$ during Hawking's evaporation
process, it will always remain unaffected. That is, unless some
external forces apply to remove the cosmological constant
$\Lambda^*$ from the spacetime geometries, it will continue to
exist along with the electrically radiating objects, rotating or
non-rotating forever. This leads to the proof of the theorem 6
cited above for  the rotating black holes. The metric (3.71)
can be written in Kerr-Schild form:
\begin{eqnarray*}
g_{ab}^{\rm NMdS}=g_{ab}^{\rm dS} +2Q(u,r,\theta)\ell_a\ell_b
\end{eqnarray*}
where $Q(u,r,\theta) = \{r{\cal M}(u)+e^2(u)/2\}\,R^{-2}$, where
$g_{ab}^{\rm dS}$ is the rotating de Sitter cosmological metric.

\vspace*{0.2in}

{\sl (iv) Variable-charged rotating Vaidya-Bonnor-monopole Black hole} \\\

Here, we shall incorporate the Hawking radiation, how the rotating
variable-charged black hole is affected in the classical
space-time metric when the electric charge $e$ is taken as a
function of $u$ and $r$ in the Einstein-Maxwell field equations.
The line element for the rotating Vaidya-Bonnor black hole with
$e(u,r)$ is
\begin{eqnarray}
ds^2&=&[1-R^{-2}\{2rM(u)-e^2(u,r)+br^2\}]\,du^2+2du\,dr \cr
&+&2aR^{-2}\{2rM(u)-e^2(u,r)+br^2\}\,{\rm sin}^2
\theta\,du\,d\phi-2a\,{\rm sin}^2\theta\,dr\,d\phi \cr
&-&R^2d\theta^2-\{(r^2+a^2)^2 -\Delta^*a^2\,{\rm
sin}^2\theta\}\,R^{-2}{\rm sin}^2\theta\,d\phi^2,
\end{eqnarray}
where $\Delta^*=r^2(1-b)-2rM(u)+a^2+e^2(u,r)$. This metric will
also reduce to rotating Vaidya-Bonnor-monopole solution when
$e(u,r)=e(u)$, initially. Then the Einstein-Maxwell field
equations for the above metric with $e(u,r)$ can be solved to
obtain the affected quantities with the monopole constant $b$:
\begin{eqnarray}
 \psi_2 &=&{1\over\overline R\,\overline
R\,R^2}\Big[\Big\{-RM(u)+e^2(u,r)\Big\}-\overline
R\,e(u,r)\,e(u,r)_{,r}\cr && +{1\over 6}\,\overline R\,\overline
R\,\Big\{e(u,r)\,e(u,r)_{,r}\Big\}_{,r}-{b\over 6}\,
\Big(RR+2\,r\,a\,i{\rm cos}\theta\Big)\Big],\\
\phi_{11}&=&{1\over{4\,R^2\,R^2}}\,\Big[ 2\,e^2(u,r)
-4re(u,r)\,e(u,r)_{,r} \cr &&+R^2\{e(u,r)\,e(u,r)_{,r}\}_{,r}\Big]
+{1\over{4\,R^2\,R^2}}\,b\,(r^2-a^2\,{\rm cos}^2\theta),\cr
\Lambda &= &-{1\over
{12\,R^2}}\,\Big\{e^2(u,r)_{,r}+e(u,r)\,e(u,r)_{,r\,r}\Big\}+{b\over{12\,R^2}},\
\end{eqnarray}
where $\psi_3$ and $\psi_4$ are same as in (3.20) of case $3(ii)$.
According to the total energy momentum tensor, we shall,
without loss of generality, have the following decompositions for
Vaidya-Bonnor-monopole:
\begin{eqnarray}
\phi^{\rm (E)}_{11}&=&{1\over{4\,R^2\,R^2}}\,\Big[ 2\,e^2(u,r)
-4re(u,r)\,e(u,r)_{,r} \cr
&&+R^2\{e(u,r)\,e(u,r)_{,r}\}_{,r}\Big],\\
\phi^{\rm (m)}_{11}&=&{1\over{4\,R^2\,R^2}}\,b\,(r^2-a^2\,{\rm
cos}^2\theta),\cr \Lambda^{\rm (E)} &= &-{1\over
{12\,R^2}}\,\Big\{e^2(u,r)_{,r}+e(u,r)\,e(u,r)_{,r\,r}\Big\},\cr
\Lambda^{\rm (m)} &=&{b\over{12\,R^2}}.
\end{eqnarray}
For electromagnetic field, the Ricci scalar $\Lambda^{\rm (E)}$
given above must vanish. This yields
\begin{equation} e^2(u,r) = 2\,rm_1(u)\ + C(u)
\end{equation}
where $m_1(u)$ and $C(u)$ are real functions of u. Then, using
this result in equation (3.77) we obtain the Ricci scalar
\begin{equation}
\phi^{\rm (E)}_{11}={C(u)\over 2\,R^2\,R^2}.
\end{equation}
Accordingly, the Maxwell scalar may become, after identifying the
function $C(u)=e^2(u)$,
\begin{equation}
\phi_1={e(u)\over\surd 2\,\overline R\,\overline R}.
\end{equation}
Then, using the relation (3.79) in (3.75), we find the changed
Weyl scalar $\psi_2$
\begin{eqnarray}
 \psi_2&=&{1\over{\overline R\,\overline
R\,R^2}}\Big[-\{M(u)-m_1(u)\}R + e^2(u)-{b\over 6}\,\Big(RR+2\,r\,a\,i{\rm
cos}\theta\Big)\Big],\\
\end{eqnarray}
but changes in $\psi_3$ and $\psi_4$ are same as (3.28) of case
3(ii). There is no change in monopole constant $b$.
Thus, we have the line element with the change of mass as
\begin{eqnarray}
ds^2&=&\Big[1-R^{-2}\,\Big\{2r\Big(M(u)-m_1(u)\Big)-e^2(u)+br^2\Big\}\Big]\,du^2+2du\,dr
\cr &+&2aR^{-2}\{2r\Big(M(u)-m_1(u)\Big)-e^2(u)+br^2\}\,{\rm
sin}^2 \theta\,du\,d\phi-2a\,{\rm sin}^2\theta\,dr\,d\phi \cr
&-&R^2d\theta^2-\{(r^2+a^2)^2 -\Delta^*a^2\,{\rm
sin}^2\theta\}\,R^{-2}{\rm sin}^2\theta\,d\phi^2,
\end{eqnarray}
where $\Delta^*=r^2(1-b)-2r\{M(u)-m_1(u)\}+a^2+e^2(u)$. This
introduction of real function $m_1(u)$ in the metric (3.84)
suggests that the first electrical radiation of rotating black
hole may reduce the original gravitational mass $M(u)$ by a
quantity $m_1(u)$. If one considers another radiation by taking
$e(u)$ in (3.84) to be a function of $u$ and $r$ with the mass
$M(u)-m_1(u)$, then the Einstein-Maxwell field equations yield to
reduce this mass by another real function $m_2(u)$; i.e., after
the second radiation, the mass may become
$M(u)-\{m_1(u)+m_2(u)\}$. Here again, the Maxwell scalar $\phi_1$
remains the same form after the second radiation also. Thus, if
one consider the $n$-time radiations taking every time the charge
$e(u)$ to be function of $u$, the Maxwell scalar $\phi_1$ will
remains the same. Taking Hawking's radiation of black holes, one
might expect that the total mass of black hole may be radiated
away just leaving ${\cal M}(u)$ equivalent to Planck mass of about
$10^{-5}g$, that is, $M(u)$ may not be exactly equal to $m_1(u) +
m_2(u) + m_3(u) + m_4(u) + . . .+ m_n(u)$, but has a difference of
about Planck-size mass, as in the case of non-rotating black hole.
Otherwise, the total mass of black hole may be evaporated
completely after continuous radiation when ${\cal M}(u) = 0$, that
is, $M(u) = m_1(u) + m_2(u) + m_3(u) + m_4(u) + . . .+ m_n(u)$.
Here one may regard that the rotating variable-charged black hole
might be radiated completely away all its mass just leaving the
electrical charge $e(u)$ and monopole constant $b$ only.
One could observe this situation in the form of classical space-time
metric as
\begin{eqnarray}
ds^2&=&[1+R^{-2}\{e^2(u)-br^2\}]\,du^2+2du\,dr \cr
&-&2aR^{-2}\{e^2(u)-br^2\}\,{\rm sin}^2 \theta\,du\,d\phi-2a\,{\rm
sin}^2\theta\,dr\,d\phi \cr &-&R^2d\theta^2-\{(r^2+a^2)^2
-\Delta^*a^2\,{\rm sin}^2\theta\}\,R^{-2}{\rm
sin}^2\theta\,d\phi^2,
\end{eqnarray}
with the charge $e(u)$ and the monopole constant $b$,
but no mass, where $\Delta^*=r^2\,(1-b)+a^2+e^2(u)$. The metric
(3.85) will describe a rotating `instantaneous' charged monopole black hole
as the remnant of Vaidya-Bonnor-monopole space,
otherwise the metric describes the rotating charged monopole black
hole with the horizons at $r_{\pm}=\pm{1\over{1-b}}\,[
\surd{(b-1)\,\{a^2+e^2(u)}\}]$. Hence one
may refer to the solution (3.85) as an 'instantaneous' charged
black hole with the surface gravity,
\begin{eqnarray}
{\cal
K}={-1\over{r_{+}\,R^2}}\,\Big[r_{+}\,\Big\{r_{+}\,\Big(1-{b\over
2}\Big)\Big\}+{e^2(u)\over 2}\Big].
\end{eqnarray}
The entropy and angular velocity of the horizon are respectively
obtained by
 \begin{eqnarray}
{\cal S}=\pi\,(r^2_{+}+a^2),\;\;\; {\rm and}\;\;\; \Omega_{\rm H}
={-a\,\{e^2(u)-b\,r^2\}\over(r^2+a^2)^2}\Big|_{r=r_{+}}.
\end{eqnarray}
At this stage, the
Weyl scalar $\psi_2$ takes the form
\begin{eqnarray}
\psi_2={1\over{\overline R\,\overline R\,R^2}}\Big\{
e^2(u)-{b\over6}\,\Big(R\,R+2\,r\,a\,i\,{\rm
cos}\theta\Big)\Big\},
\end{eqnarray}
and $\psi_3$, $\psi_4$ are remained same as in (3.34) and (3.35)
showing the gravity on the surface of the remaining solution
depending on the electric charge $e(u)$ and the monopole charge
$b$ coupling with the rotational parameter $a$; however, the
Maxwell scalar $\phi_1$ and the Ricci scalar $\Lambda^{(\rm m)}$
remain the same as in (3.81) and (3.78) respectively. This
completes the other part of the theorem 3 cited above for the case
embedded into the monopole universe. It means that there may be
rotating black holes in the universe whose masses are completely
radiated; their gravity depends on the electric charge of the body
and the monopole charge $b$, and their metrics appear similar to
that in (3.85). Here the idea of this complete evaporation of
masses of radiating black holes embedded in the monopole space is
in agreement with that of Hawking's evaporation of black holes. Here,
we again consider the charge $e(u)$ to be function of $u$ and $r$ for
next radiation in (3.85), so that from the Einstein's field equations,
we get the scalar $\Lambda^{(\rm E)}$ as given above and the same
scalar $\Lambda^{(\rm m)}$ as in (3.78). Then the vanishing of this
$\Lambda^{(\rm E)}$ for electromagnetic field will lead to create a
new mass (say $m^*_1(u)$) in the remaining space-time geometry (3.85).
For the second radiation, we again consider the charge $e(u)$ to be
function of $u$ and $r$ in the field equations with the mass
$m^*_1(u)$. Then the vanishing of $\Lambda^{(\rm E)}$ will lead to
increase the new mass by another quantity $m^*_2(u)$ (say) i.e.,
after the second radiation of (3.85) the new mass will be
$m^*_1(u) +m^*_2(u)$. If this radiation process continues further
for a long time, the new mass will increase gradually as
\begin{equation}
{\cal M}^*(u)=m^*_1(u) + m^*_2(u) + m^*_3(u) + m^*_4(u) + . . .
.\,\,
\end{equation}
Then, the spacetime metric will take the form
\begin{eqnarray}
ds^2&=&\Big[1+R^{-2}\Big\{2r{\cal
M}^*(u)+e^2(u)-b\,r^2\Big]\,du^2+2du\,dr \cr
&&-2aR^{-2}\Big\{2r{\cal M}^*(u)+e^2(u)-b\,r^2\Big\}\,{\rm sin}^2
\theta\,du\,d\phi -2a\,{\rm sin}^2\theta\,dr\,d\phi \cr
&&-R^2d\theta^2-\Big\{(r^2+a^2)^2 -\Delta^*a^2\,{\rm
sin}^2\theta\Big\}\, R^{-2}{\rm sin}^2\theta\,d\phi^2.
\end{eqnarray}
where $\Delta^*=r^2\,(1-b)+2r{\cal M}^*(u)+a^2+e^2(u)$. The
changed NP quantities $\psi_2$, $\psi_3$ and $\psi_4$ are as
follows
\begin{eqnarray}
\psi_2={1\over{\overline R\,\overline R\,R^2}}\,\Big\{{\cal
M}^*(u)R+e^2(u)-{b\over6}\,\Big(R\,R+2\,r\,a\,i\,{\rm
cos}\theta\}\Big)\Big\},
\end{eqnarray}
and $\psi_3$, $\psi_4$  are given in (3.43) and $\phi_1$ and
$\Lambda^{(\rm m)}$ are remained unchanged. Here we find that
the classical spacetime (3.90) describes a rotating negative mass
naked singularity embedded into the monopole space.
Thus, from the above it follows the proof of the rotating part of
theorem 5 that {\sl during the radiation process, after the
complete evaporation of masses of variable-charged non-stationary
embedded black holes, the electrical radiation will continue
indefinitely creating embedded negative mass naked singularities}.
 We have also shown the possible changes in the mass of
the rotating variable charged Vaidya-Bonnor-monopole black hole
without affecting the Maxwell scalar $\phi_1$ as well as the
monopole constant $b$, and accordingly, metrics are cited for
future use. Thus, this completes the proof of other part of the
theorem 1 for the embedded rotating Vaidya-Bonnor-monopole black
hole. Also since there is no effect on the monopole constant $b$
during Hawking's evaporation process, it will always remain
unaffected. That is, unless some external forces apply to remove
the monopole constant $b$ from the spacetime geometries, it will
continue to exist along with the electrically radiating rotating
objects forever. This leads to the proof of the theorem 6 for
rotating embedded black holes that {\sl if an electrically radiating
non-stationary black hole is embedded into a space, it will continue
to embed into the same space forever}. The metric (3.90) can be
expressed in Kerr-Schild form on the monopole background
\begin{eqnarray*}
g_{ab}^{\rm NMm}=g_{ab}^{\rm m} +2Q(u,r,\theta)\ell_a\ell_b
\end{eqnarray*}
where $Q(u,r,\theta) = \{r{\cal M}(u)+e^2(u)/2\}\,R^{-2}$. Here,
$g_{ab}^{\rm m}$ is the rotating monopole metric and $\ell_a$ is
geodesic, shear free, expanding and non-zero twist null vector for
both $g_{ab}^{\rm m}$ as well as $g_{ab}^{\rm NMm}$.

\vspace*{0.2in}

{\sl (v) Variable charged rotating Vaidya-Bonnor-Kerr black
hole} \\\

Here, one may incorporate the Hawking radiation, how the rotating
variable-charged black hole affect in the classical space-time
metric when the electric charge $e$ is taken as a function of $u$
and $r$ in the Einstein-Maxwell field equations. The line element
rotating Vaidya-Bonnor-Kerr black hole with $e(u,r)$ is
\begin{eqnarray}
d
s^2&=&\Big[1-R^{-2}\Big\{2r\Big(\tilde{m}+M(u)\Big)
-e^2(u,r)\Big\}\Big]\,du^2+2du\,dr
\cr &&+2aR^{-2}\Big\{2r\Big(\tilde{m}+M(u)\Big)-e^2(u,r)\Big\}{\rm
sin}^2\theta\,du\,d\phi \cr && -2a{\rm sin}^2\theta\,drd\phi
-R^2d\theta^2-\Big\{(r^2+a^2)^2 -\Delta^*a^2{\rm
sin}^2\theta\Big\}R^{-2}{\rm sin}^2\theta\,d\phi^2,
\end{eqnarray}
where $\Delta^*=r^2-2r{\tilde{m}+M(u)}+a^2+e^2(u,r)$. This metric
will also reduce to rotating Vaidya-Bonnor-Kerr solution when
$e(u,r)=e(u)$ initially, the Einstein-Maxwell field equations for
the above metric with $e(u,r)$ can be solved to obtain the
following quantities:
\begin{eqnarray}
\Lambda &= &-{1\over
{12\,R^2}}\,\Big\{e^2(u,r)_{,r}+e(u,r)\,e(u,r)_{,rr}\Big\},\\
\psi_2&=&{1\over\overline R\,\overline
R\,R^2}\Big[-R\Big\{\tilde{m}+M(u)\Big\}+e^2(u,r) +{\overline
R\,\overline R\over 6}\Big\{e(u,r)\,e(u,r)_{,r}\Big\}_{,r} ,\cr
&&-\overline R\Big\{e(u,r)\,e(u,r)_{,r}\Big\}\Big],
\end{eqnarray}
where $\psi_3$ and $\psi_4$ are same as (3.20) and $\phi_{11}$,
$\phi_{12}$ and $\phi_{22}$ are affected as in (3.21).
The scalar $\Lambda=\Lambda^{\rm E}$ must vanish for this
rotating metric. Thus, vanishing $\Lambda^{\rm E}$ of the
equation (3.93) implies that
\begin{equation}
e^2(u,r) = 2\,rm_1(u)\ + C(u)
\end{equation}
where $m_1(u)$ and $C(u)$ are real functions of $u$. Then, using
this result, we obtain the Ricci scalar
\begin{equation}
\phi_{11}={C(u)\over 2\,R^2\,R^2}.
\end{equation}
Accordingly, the Maxwell scalar may become, after identifying the
function $C(u)=e^2(u)$,
\begin{equation}
\phi_1={e(u)\over\surd 2\,\overline R\,\overline R}.
\end{equation}
Then using the relation (3.95) in (3.94), we find the changed Weyl
scalar
\begin{eqnarray}
\psi_2={1\over\overline R\,\overline
R\,R^2}\Big[-R\Big\{\tilde{m}+M(u)-m_1(u)\Big\}+e^2(u)\Big],
\end{eqnarray}
and changes in  Weyl scalars $\psi_3$ and $\psi_4$ are same as
(3.28) of case 3(ii). Thus, we have the line element with the
change of mass as
\begin{eqnarray}
d s^2&=&\Big[1-R^{-2}\Big\{2r\Big(\tilde{m}+M(u)-m_1(u)\Big)-
e^2(u)\Big\}\Big]\,du^2+2du\,dr \cr
&&+2aR^{-2}\Big\{2r\Big(\tilde{m}+M(u)-m_1(u)\Big)-e^2(u)\Big\}\,{\rm
sin}^2 \theta\,du\,d\phi \cr && -2a\,{\rm sin}^2\theta dr d\phi
-R^2d\theta^2-\Big\{(r^2+a^2)^2 -\Delta^*a^2{\rm
sin}^2\theta\Big\}R^{-2}{\rm sin}^2\theta d\phi^2
\end{eqnarray}
where $\Delta^*=r^2-2r\{\tilde{m}+M(u)-m_1(u)\}+a^2+e^2(u)$. This
introduction of real function $m_1(u)$ in the metric (3.99)
suggests that the first electrical radiation of rotating black
hole may reduce the original gravitational mass $M(u)$ by a
quantity $m_1(u)$. If one considers another radiation by taking
$e(u)$ in (3.99) to be a function of $u$ and $r$ with the mass
$M(u)-m_1(u)+\tilde{m}$, then the Einstein-Maxwell field equations
yield to reduce this mass by another real function $m_2(u)$; i.e.,
after the second radiation, the mass may become
$M(u)-\{m_1(u)+m_2(u)\}+\tilde{m}$. Here again, the Maxwell scalar
$\phi_1$ remains the same form after the second radiation also.
Thus, if one considers the $n$-time radiations taking every time
the charge $e(u)$ to be function of $u$, the Maxwell scalar
$\phi_1$ will be the same. Taking Hawking's radiation of black
holes, one might expect that the total mass of black hole may be
radiated away just leaving ${\cal M}(u)$ equivalent to Planck mass
of about $10^{-5}g$ and $m$ remain same. that is, $M(u)$ may not
be exactly equal to $m_1(u) + m_2(u) + m_3(u) + m_4(u) + . . .+
m_n(u)$, but has a difference of about Planck-size mass, as in the
case of non-rotating black hole. Otherwise, the total mass of
black hole may be evaporated completely after continuous radiation
when $M(u) = m_1(u) + m_2(u) + m_3(u) + m_4(u) + . . .+ m_n(u)$,
just leaving the mass $m$ and electrical charge $e(u)$ only. Thus,
one could observe this situation in the form of classical
space-time metric as
\begin{eqnarray}
d
s^2&=&\Big[1-R^{-2}\Big\{2r\,\tilde{m}-e^2(u)\Big\}\Big]\,du^2+2du\,dr
\cr &&+2aR^{-2}\Big\{2r\,\tilde{m}-e^2(u)\Big\}\,{\rm sin}^2
\theta\,du\,d\phi \cr && -2a\,{\rm sin}^2\theta\,dr\,d\phi
-R^2d\theta^2 \cr && -\Big\{(r^2+a^2)^2 -\Delta^*a^2\,{\rm
sin}^2\theta\Big\}\,R^{-2}{\rm sin}^2\theta\,d\phi^2,
\end{eqnarray}
where $ \Delta^*=r^2-2r\,\tilde{m}+a^2+e^2(u)$. The metric (3.100)
will describe a rotating `instantaneous' charged black hole, i.e.
a rotating charged Kerr black hole with $\tilde{m}>
a^2+e^2(u)$.
At this stage, the Weyl scalar $\psi_2$ takes the form
\begin{eqnarray}
\psi_2={1\over\overline R\,\overline
R\,R^2}\Big\{-R\,\hat{m}+e^2(u)\Big\},
\end{eqnarray}
and $\psi_3$, $\psi_4$ are remain same as (3.34) and (3.35). The
surface gravity of the horizon at
$r=r_{+}=\tilde{m}+\surd[\tilde{m}^2-\{a^2+e^2(u)\}]$, is
\begin{eqnarray}
{\cal
K}=-\frac{1}{r_{+}\,R^2}\,\Big\{r_{+}\,\sqrt{\tilde{m}^2-a^2-e^2(u)}
+\frac{e^2(u)}{2}\Big\}.
\end{eqnarray}
Then we find the entropy and angular velocity of the horizon
respectively as follows:
\begin{eqnarray}
{\cal
S}=2\pi\,\tilde{m}\,\{\tilde{m}+\sqrt{\tilde{m}^2-a^2-e^2(u)}\}-e^2(u)\;\;\;
{\rm and}\;\;\; \Omega_{\rm H}
={a\,\{2\,r\,\tilde{m}-e^2(u)\}\over(r^2+a^2)^2}\Big|_{r=r_{+}},
\end{eqnarray}
showing the gravity on the surface of the remaining solution
depending on the electric charge $e(u)$ and $\tilde{m}$; however,
the Maxwell scalar $\phi_1$ remains the same as in (3.97). Here,
one may consider again the charge $e(u)$ to be function
of $u$ and $r$ for next radiation in (3.100), so that one must get
from the Einstein's field equations the scalar $\Lambda$ as given
in equation (3.93). Then the vanishing of this $\Lambda$ for
electromagnetic field, there may be creation of a new mass (say
$m^*_1(u)$ in the remaining space-time geometry. If this radiation
process continues forever, the new mass may increase gradually as
\begin{equation}
{\cal M}^*(u)=m^*_1(u) + m^*_2(u) + m^*_3 (u)+ m^*_4(u) + . . .
.\,\,
\end{equation}
However, it appears that this new mass would never decrease. Then
the space-time geometry may take the form
\begin{eqnarray}
ds^2&=&\Big[1+R^{-2}\Big\{2r\Big({\cal
M}^*(u)-\tilde{m}\Big)+e^2(u)\Big\}\Big]\,du^2+2du\,dr \cr
&-&2aR^{-2}\Big\{2r\Big({\cal
M}^*(u)-\tilde{m}\Big)+e^2(u)\Big\}\,{\rm sin}^2
\theta\,du\,d\phi-2a\,{\rm sin}^2\theta\,dr\,d\phi \cr
&-&R^2d\theta^2-\{(r^2+a^2)^2 -\Delta^*a^2\,{\rm
sin}^2\theta\}\,R^{-2}{\rm sin}^2\theta\,d\phi^2,
\end{eqnarray}
where, $\Delta^*=r^2+2r\Big\{{\cal
M}^*(u)-\tilde{m}\Big\}+a^2+e^2(u)$. The Weyl scalar $\psi_2$ and
other NP coefficients are calculated from the Einstein-Maxwell
field equations as
\begin{eqnarray}
\psi_2={1\over{\overline R\,\overline R\,R^2}}\,\Big[R\Big\{{\cal
M}^*(u)-\tilde{m}\Big\}+e^2(u)\Big],
\end{eqnarray}
with $\psi_3$, $\psi_4$, given in (3.43) and $\phi_1$ and
$\Lambda^{\rm (E)}$ are remain unchanged. The metric (3.105)
may be regarded to describe a rotating negative mass naked
singularity embedded into Kerr black hole. We have
presented the possible changes in the mass of the rotating charged
black hole without affecting the Maxwell scalar $\phi_1$ and
Kerr mass accordingly, metrics are cited for future use. Thus, this
completes the proofs of other parts of the theorem 5 that
{\sl the electrical radiation will continue indefinitely creating
embedded negative mass naked singularities}. The metric (3.105)
can be written in Kerr-Schild form on the Kerr background as
\begin{equation}
g_{ab}^{\rm NMK}=g_{ab}^{\rm K} +2Q(u,r,\theta)\ell_a\ell_b
\end{equation}
where $Q(u,r,\theta) = \{r{\cal M}(u)+e^2(u)/2\}R^{-2}$. Here, $g_{ab}^{\rm
K}$ is the rotating Kerr metric and $\ell_a$ is geodesic, shear
free, expanding and non-zero twist null vector for both
$g_{ab}^{\rm K}$ as well as $g_{ab}^{\rm NMK}$.

 \vspace*{0.12in}

\begin{center}
{\bf 3. Conclusion}
\end{center}
\setcounter{equation}{0}
\renewcommand{\theequation}{3.\arabic{equation}}

In this paper, we have presented NP quantities for a rotating
spherically symmetric metric with two variables $u, r$ in the appendix
(A1) below. With the help of these NP quantities, we have
first derived a class of non-stationary rotating solutions including
Vaidya-Bonnor-de Sitter, Vaidya-Bonnor-monopole, and Vaidya-Bonnor-Kerr.
Then we studied the gravitational structure of the solutions by
observing the nature of the energy momentum tensors of respective
spacetime metrics. These solutions describe embedded rotating black
holes. For example, Vaidya-Bonnor black hole is embedded
into the rotating de Sitter cosmological space to produce
the Vaidya-Bonnor-de Sitter cosmological black hole and similarly,
Vaidya-Bonnor-monopole and Vaidya-Bonnor-Kerr black holes. The embedded
rotating solutions have also been expressed in terms of
Kerr-Schild ansatze in order to indicate them as solutions of
Einstein's field equations. These ansatze show the extensions of
those of Glass and Krisch [24] and Xanthopoulos [25]. This completes
the proof of the theorem 7 that {\sl every embedded black hole,
stationary or non-stationary, is expressible
in Kerr-Schild ansatz}. The theorem 7 is valid for both the
{\sl stationary} [26] and {\sl non-stationary} embedded black holes here.

The remarkable feature of the analysis  of rotating solutions in
this paper is that all the rotating solutions are non-stationary
algebraically special of the Petrov classification of spacetime
metric, possess the same null vector $\ell_a$, given in(A2),
which is geodesic, shear free, expanding as well as non-zero
twist (A7). From the study of the rotating solutions
we find that some solutions after making rotation have disturbed
their gravitational structures. For example, after making rotation,
the Vaidya metric with $e(u)= \Lambda^*=0$ in (2.19) becomes
algebraically special in Petrov classification with the non-vanishing
$\psi_2$, $\psi_3$ and $\psi_4$ (2.22), and a null vector $\ell_a$
which is geodesic, shear free, expanding and non-zero twist.
Similarly, the rotating de Sitter solution (2.3) with
$M(u)=e(u)=0$ becomes
Petrov type D spacetime metric, where the rotating parameter
$a$ is coupled with the cosmological constant in the expressions
of $\phi_{11}=-(1/2\,R^2\,R^2)\Lambda^*r^2a^2\,{\rm cos}^2\theta,$
and $\psi_2=(1/3\overline R\,\overline R\,R^2)\Lambda^*r^2a^2{\rm
cos}^2\theta$. The rotating monopole solution (2.18) with
$M(u)=e(u)=0$ possesses the energy momentum tensor
with the monopole pressure $p$, where the monopole charge $b$
couples with the rotating parameter $a$ as
in $p=(-1/K\,R^2\,R^2)\,b\,a^2{\rm cos}^2\theta$.
The method adopted here with Wang-Wu functions might be
another possible version for obtaining {\sl non-stationary}
rotating black hole solutions with visible energy momentum tensors
describing the interaction of different matter fields with
well-defined physical properties like Guth's modification of
$T_{ab}$ (2.14). It is believed that such interactions of
different matter fields as in (2.9), (2.14) and (2.26)
have not seen published before. We have also found the direct
involvement of the rotation parameter $a$ in each expression
of the surface gravity and the angular velocity, which shows
the important of the study of non-stationary rotating embedded,
black holes in order to understand the nature of different black
holes located in the universe.

In section 3, we find that the changes in the masses of
non-embedded as well as embedded black holes take place due to
the vanishing of Ricci scalar of
electromagnetic fields with the charge $e(u,r)$. It is also shown
that the Hawking's radiation can be expressed in classical
spacetime metrics, by considering the charge $e(u)$ to be a
function of $u$, and $r$ of Vaidya-Bonnor-de Sitter,
Vaidya-Bonnor-monopole and Vaidy-Bonnor-Kerr black holes.
That is, every electrical radiation produces a change in the mass
of the {\sl non-stationary} charged black holes. It may be
concluded the proof of the theorem 1. These changes in the mass
of black holes, non-embedded [non-rotating (3.1) and rotating (3.18)]
and embedded into de Sitter, monopole and Kerr spaces, after
every electrical radiation, describe the relativistic aspect of
Hawking's evaporation of masses of black holes in the classical
spacetime metrics.  Thus, we find that the black hole evaporation
process is due to the electrical radiation of the variable charge
$e(u, r)$ in the energy momentum tensor describing the change in the
mass in classical spacetime metrics which is in agreement with
Boulware's suggestion [7]. The Hawking's evaporation of masses and
the creation of non-embedded negative mass naked singularities
(3.16), (3.40), and embedded ones (3.71), (3.90) and (3.105), are
also due to the continuous electrical radiation. This suggests
that, if one accepts the continuous electrical radiation to lead
the complete evaporation of the original non-stationary mass
of black holes, then the same radiation will also lead to
the creation of new non-stationary mass to
form negative mass naked singularities. This clearly indicates
that an electrically radiating  embedded black hole will not
disappear completely, which is against the suggestion made in
[2, 3, 5]. It is noted that we observe the different results from
the study of {\sl embedded} and {\sl non-embedded} black holes.
In the embedded cases here above, the presence of other quantities
like the cosmological constant $\Lambda^{*}$ (3.66), the monopole
constant $b$ (3.85) and the Kerr mass $\tilde{m}$ (3.100)
presumably prevent the
disappearance of embedded radiating black holes during the
radiation process, and thereby, the formation of `instantaneous'
rotating charged black holes (3.66), (3.85) and (3.100).
In {\sl non-embedded} cases (3.15) and (3.32), the
disappearance of a black hole during radiation process is
unavoidable, however occurs for an instant with the formation of
`instantaneous' naked singularity with zero mass, before
continuing its next radiation.

It appears that (i) the changes in the mass of black holes, (ii)
the formation of `instantaneous' naked singularities  with zero
mass and (iii) the creation of `negative mass naked singularities'
in {\sl non-embedded}, non-rotating as well as rotating,
Vaidya-Bonnor black holes, (3.16) and (3.40) are presumably
the correct formulation in
non-stationary classical spacetime metrics of the three
possibilities of black hole evaporation suggested by Hawking
and Israel [3]. However,
the creation of `negative mass naked singularities' may be a
violation of Penrose's cosmic censorship hypothesis [9]. It is
found that (i) the changes in the masses of embedded black holes,
(ii) the formation of `instantaneous' rotating charged black
holes (3.66), (3.85) and (3.100), and (iii) the creation of
embedded `negative mass naked singularities' in Vaidya-Bonnor-de
Sitter, Vaidya-Bonnor-monopole and Vaidya-Bonnor-Kerr black holes
might be the mathematical formulations in non-stationary
classical spacetime metrics of the three possibilities of black hole
evaporation [3]. All embedded black holes discussed here can be
expressed in Kerr-Schild ansatze, accordingly their consequent
negative mass naked singularities are also expressible in
Kerr-Schild forms showing them as solutions of Einstein's field
equations. It is also observed that once a charged black hole is
embedded into some spaces, it will continue to embed forever
through out its Hawking evaporation process. For example,
Vaidya-Bonnor black hole is embedded into the rotating de Sitter
universe, it continues to embed as `instantaneous'
charged black hole in (3.66) and embedded negative mass naked
singularity as in (3.71). There Hawking's radiation does not
affect the cosmological constant $\Lambda^{*}$ through out
the evaporation process of Vaidya-Bonnor mass, and similarly,
in the cases of Vaidya-Bonnor-monopole as well as
Vaidya-Bonnor-Kerr black holes we find that
the monopole constant and Kerr mass remain unaffected. This
means that the embedded negative mass naked singularities
(3.71), (3.90) and (3.105) possess the total energy momentum
tensors (2.9), (2.26) and (2.40) respectively with mass
${\cal M}^{*}(u)$, as the change
in the mass $M(u)$, due to continuous radiation, affects
the energy momentum tensors. Thus, it may be concluded that
once a black hole is embedded
into some universe, it will continue to embed forever without
disturbing the nature of matters of the back ground spaces.
This completes the proof of theorem 6. If one accepts the
Hawking continuous evaporation of charged black holes, the loss of
mass and creation of new mass are the process of the continuous
radiation. So, it may also be concluded that once electrical
radiation starts, it will continue to radiate forever describing
the various stages of the life of radiating black holes.

Also, we find from the above that the change in the mass of black
holes, {\sl embedded} or {\sl non-embedded}, takes place due to
the Maxwell scalar ${\phi_1}$, remaining unchanged in the field
equations during continuous radiation. So, if the Maxwell scalar
${\phi_1}$ is absent from the space-time geometry, there will be
no radiation, and consequently, there will be no change in the
mass of the non-stationary black holes. Therefore, we cannot, theoretically,
expect to observe  such {\sl relativistic change} in the mass of
{\sl uncharged}, non-rotating as well as rotating, Vaidya black holes.
It is observed that the non-stationary classical spacetime
metrics discussed above would describe the possible life style of
radiating embedded black holes at different stages during their
continuous radiation. These {\sl embedded} non-stationary
classical spacetimes metrics
describing the changing life style of black holes are different from the
{\sl non-embedded} non-stationary ones in various respects shown
above. Here the study of these embedded solutions suggests the
possibility that in an early universe there might be some
non-stationary black holes, which might have embedded into
some other spaces possessing
different matter fields with well-defined physical properties.

\vspace*{.2in}

\begin{appendix}
\setcounter{equation}{0}
\renewcommand{\theequation}{A\arabic{equation}}
\begin{center}
{\bf Appendix}
\end{center}

In this appendix we present a rotating metric with the function
$M(u,r)$. The line element will be of the form [13]
\begin{eqnarray}
d s^2&=&\{1-2rM(u,r)R^{-2}\}\,du^2+2du\,dr +4arM(u,r)R^{-2}\,{\rm
sin}^2 \theta\,du\,d\phi \cr &&-2a\,{\rm sin}^2\theta\,dr\,d\phi
-R^2d\theta^2 -\{(r^2+a^2)^2 -\Delta^*a^2\,{\rm sin}^2\theta\}\,
R^{-2}{\rm sin}^2\theta\,d\phi^2,
\end{eqnarray}
where $R=r+ia\,{\rm cos}\theta$ and $\Delta^*=r^2-2rM(u,r)+a^2$.
Then the covariant complex null tetrad vectors for the metric can
be chosen as follows
\begin{eqnarray}
&&\ell_a=\delta^1_a -a\,{\rm sin}^2\theta\,\delta^4_a,\\
&&n_a=\frac{\Delta^{*}}{2\,R^2}\,\delta^1_a+ \delta^2_a
-\frac{\Delta^{*}}{2\,R^2}\,\,a\,{\rm sin}^2\theta\,\delta^4_a,\cr
&&m_a=-{1\over\surd 2R}\,\Big\{-ia\,{\rm
sin}\,\theta\,\delta^1_a+R^2\,\delta^3_a +i(r^2+a^2)\,{\rm
sin}\,\theta\,\delta^4_a\Big\}, \\
&&\overline m_a=-{1\over\surd 2\overline R}\,\Big\{ia\,{\rm
sin}\,\theta\,\delta^1_a +R^2\,\delta^3_a-i(r^2+a^2)\,{\rm
sin}\,\theta\,\delta^4_a\Big\}.\nonumber
\end{eqnarray}
Then the Newman-Penrose spin-coefficients, the Ricci scalars and
Weyl scalars for the metric (A1) are given bellow:
\begin{eqnarray}
&&\kappa=\sigma=\lambda=\epsilon=0, \cr &&\rho=-{1\over\overline
R},\;\;\;\; \mu=-\frac{\Delta^{*}}{2\overline R\,R^2},\cr
&&\alpha={(2ai-R\,{\rm cos}\,\theta)\over{2\surd 2\overline
R\,\overline R\,{\rm sin}\,\theta}},\;\;\; \beta={{\rm
cot}\,\theta\over2\surd 2R}, \cr &&\pi={i\,a\,{\rm
sin}\,\theta\over{\surd 2\overline R\,\overline R}},\;\;\;
\tau=-{i\,a\,{\rm sin}\,\theta\over{\surd 2R^2}},\cr
&&\gamma={1\over{\surd 2\,\overline R\,R^2}}\, \Big [(r - M -
r\,M_{,r})\,\overline R - \Delta^*\Big ],\cr &&\nu={1\over{\surd
2\,\overline R\,R^2}}\,i\,a\,r\,{\rm
sin}\,\theta\,M_{,u},  \\
&&\phi_{00}=\phi_{01}=\phi_{10}=\phi_{20}=\phi_{02}=0, \cr
&&\phi_{11}={1\over{4\,R^2\,R^2}}\,\Big[4r^2M_{,r}
+R^2(-2M_{,r}-r\,M_{,r\,r})\Big],\cr &&\phi_{12}={1\over{2\surd
2R^2R^2}}\,\Big[i\,a\,{\rm sin}\,\theta \{RM_{,u}-rM_{,r}\overline
R\}\Big], \cr &&\phi_{21}={-1\over{2\surd 2R^2R^2}}\Big[i\,a\,{\rm
sin}\,\theta\{\overline RM_{,u}-rM_{,r}R\}\Big], \cr
&&\phi_{22}=-{1\over{2\,R^2\,R^2}}\,\Big [ 2r^2M_{,u}+
a^2r\,M\,{\rm sin}^2\theta\Big], \cr &&\Lambda
={1\over{12\,R^2}}\,\Big(2M_{,r}+
r\,M_{,r\,r}\Big),      \\
&&\psi_0 = \psi_1= 0,\cr &&\psi_2={1\over\overline R\,\overline
R\,R^2}\Big\{-R\,M+{\overline R\over 6}\,M_{,r}(4r+2\,i\,a\,{\rm
cos}\theta) -{r\over 6}\,\overline R\,\overline R\,M_{,rr}\Big\},
\cr &&\psi_3=-{i\,a\,{\rm sin}\theta\over{2\surd 2\overline
R\,\overline R\,R^2}}\, \Big\{(4\,r+\overline R\,)\,M_{,u} +
r\,{\overline R}\,M_{,ur}\Big\}, \cr &&\psi_4 ={{a^2\,r\,{\rm
sin}^2\theta}\over 2\overline R\,\overline
R\,R^2\,R^2}\,\Big\{R^2\,M_{,uu}-2\,r\,M_{,u}\Big\}.
\end{eqnarray}
From these NP spin coefficients we find that the rotating metric
(A1) possesses, in general, a geodesic $(\kappa=\epsilon=0)$,
shear free $(\sigma=0)$, expanding $(\hat{\theta} \neq 0)$ and
non-zero twist $(\omega^{*2} \neq 0)$  null vector $\ell_a$ [9]
where
\begin{eqnarray}
\hat{\theta}\equiv -{1\over2}(\rho + \overline\rho) ={r\over R^2},
\;\;\;\; \omega^{*2}\equiv-{1\over4}(\rho - \overline\rho)^2
=-{{a^2\cos^2\theta}\over {R^2\,R^2}}.
\end{eqnarray}

The the energy momentum tensor will take the form
\begin{eqnarray}
T_{ab}&=& \mu^*\,\ell_a\,\ell_b + 2\,\rho^*\,\ell_{(a}\,n_{b)}
+2\,p\,m_{(a}\overline m_{b)} \cr
&&+2\,\omega\,\ell_{(a}\,\overline m_{b)}
+2\,\overline\omega\,\ell_{(a}\,m_{b)},
\end{eqnarray}
with the following quantities
\begin{eqnarray}
&&\mu^* =-{1\over K\,R^2\,R^2}\Big\{2r^2M_{,u} + a^2r\,{\rm
sin}^2\theta\,M_{,uu}\Big\}, \cr &&\rho^* =  {2\,r^2\over
K\,R^2\,R^2}\,M_{,r}, \cr &&p = -{1\over K}\,\Big\{{2\,a^2\,{\rm
cos}^2\theta \over R^2\,R^2}\,M_{,r}+{r\over
R^2}\,M_{,r\,r}\Big\}, \cr &&\omega =-{i\,a\,{\rm
sin}\,\theta\over\surd
2\,K\,R^2\,R^2}\,\Big(R\,M_{,u}-r\,\overline R\,M_{,ur}\Big).
\end{eqnarray}
these quantities have the following relations with the Ricci
scalars (A5)
\begin{eqnarray}
&&K\,\mu^* = 2\,\phi_{22},\;\;  K\,\omega = - 2\,\phi_{12},\;\;
\cr &&K\,\rho^* =  2\,\phi_{11} + 6\,\Lambda,\;\;  K\,p =
2\,\phi_{11} - 6\,\Lambda.
\end{eqnarray}

Wang and Wu [12] have expanded the metric function $M(u,r)$ for
the non-rotating solution $(a = 0)$ in the power of $r$
\begin{equation}
M(u,r)= \sum_{n=-\infty}^{+\infty} q_n(u)\,r^n,
\end{equation}
where $q_n(u)$ are arbitrary functions of $u$. They consider the
above sum as an integral when the `spectrum' index $n$ is
continuous. Here using this expression in equations (A11) we can
generate rotating metrics with $a\neq 0$ as follows:
\begin{eqnarray}
&&\mu^* =-{r\over K\,R^2\,R^2}\,\sum_{n=-\infty}^{+\infty}
\Big\{2\,q_n(u)_{,u}\,r^{n+1} +a^2{\rm
sin}^2\theta\,q_n(u)_{,uu}\,r^n\Big\}, \cr &&\rho^* ={2\,r^2\over
K\,R^2\,R^2}\,\sum_{n=-\infty}^{+\infty} n\,q_n(u)\,r^{n-1},\\
&&p = -{1\over KR^2}\sum_{n=-\infty}^{+\infty} nq_n(u)r^{n-1}
\Big\{{2a^2\,{\rm cos}^2\theta \over R^2} +(n-1)\Big\}, \cr
&&\omega ={-i\,a\,{\rm sin}\,\theta\over\surd
2\,K\,R^2\,R^2}\,\sum_{n=-\infty}^{+\infty}(R-n\overline
R)\,q_n(u)_{,u}\,r^n. \nonumber
\end{eqnarray}
and other NP quantities can also be presented with this function
$q_{n}(u)$.

According to Carter [18] and York [27], we
introduce a scalar ${\cal K}$ defined by the relation $n^b
\nabla_{b}n^a={\cal K}\,n^a$, where the null vector $n^a$ in (2.5)
above is parameterized by the coordinate $u$, such that
$d/du=n^a\nabla_{a}$. Then this scalar can, in general, be
expressed in terms of NP spin coefficient $\gamma$ (A1) as
follows:
\begin{eqnarray}
{\cal K}&=& n^b\,\nabla_{b}\,n^a\ell_a =
-(\gamma+\overline\gamma).
\end{eqnarray}
On a horizon, the scalar ${\cal K}$ is called the surface gravity
of a black hole.

For future use we shall cite the four roots of the biquadratic
equation
\begin{eqnarray}
\Delta^{*}\equiv r^2-2rM(u)-\Lambda^{*}\,r^4/3+a^2+e^2=0
\end{eqnarray}
as follows for non-zero cosmological constant $\Lambda^{*}$:
\begin{eqnarray}
\Big[r_{+}\Big]_{\pm}&=&+{\frac{1}{2}}\surd\,{\Gamma} \pm
{\frac{1}{2}} \sqrt\Big[\,\frac{4}{\Lambda^{*}}- 32^{1/3}\chi
-\frac{1}{32^{1/3}\Lambda^{*}}\,\Big\{P+\sqrt{P^2-4Q}\Big\}^{1/3}\cr
&&-\frac{12}{\Lambda^{*}\surd\,\Gamma}\,M(u)\Big], \\
\Big[r_{-}\Big]_{\pm}&=&-{\frac{1}{2}}\surd\,{\Gamma} \pm
{\frac{1}{2}} \sqrt\Big[\,\frac{4}{\Lambda^{*}}- 32^{1/3}\chi
-\frac{1}{32^{1/3}\Lambda^{*}}\,\Big\{P+\sqrt{P^2-4Q}\Big\}^{1/3}\cr
&&+\frac{12}{\Lambda^{*}\surd\,\Gamma}\,M(u)\Big]
\end{eqnarray}
where
\begin{eqnarray*}
&&P=54\Big\{18\Lambda^{*}M(u)^2-12\Lambda^{*}(a^2+e^2)-1\Big\},\;\;\:
Q=\Big\{9-36\Lambda^{*}(a^2+e^2)\Big\}^3, \cr
&&\chi=\frac{1-4\Lambda^{*}(a^2+e^2)}{\Lambda^{*}\Big\{P+\sqrt{P^2-4Q}\Big\}^{1/3}},\;\;\:
\Gamma=\frac{2}{\Lambda^{*}}+ 32^{1/3}\chi
+\frac{32^{-1/3}}{\Lambda^{*}}\Big\{P+\sqrt{P^2-4Q}\Big\}^{1/3}.
\end{eqnarray*}
The calculation of these roots has been carried out by using
`Mathematica'. The area of a horizon of black hole can be
calculated as follows [9]:
\begin{eqnarray}
{\cal A} =
\int_{0}^{\pi}\int_{0}^{2\pi}\sqrt{g_{\theta\theta}g_{\phi\phi}}
\,d\theta\,d\phi\,\Big|_{\Delta^{*}=0},
\end{eqnarray}
depending on the values of the roots of $\Delta^{*}=0$. Then the
entropy on a horizon of a black hole may be obtained from the
relation ${\cal S}= {\cal A}/4$ [28].
\end{appendix}

\vspace*{0.25in}

        {\bf Acknowledgement}: The authors acknowledge their
appreciation for hospitality received from Inter-University Centre
for Astronomy and Astrophysics (IUCAA), Pune, India during their
visit.

\end{document}